\documentclass[12pt,preprint]{aastex}

\usepackage{natbib}

\begin{document}

\title{Multi-Temperature Blackbody Spectra of Thin Accretion Disks With 
and Without a Zero-Torque Inner Boundary Condition}
\author{E. R. Zimmerman, R. Narayan, J. E. McClintock, and J. M. Miller}
\affil{Harvard-Smithsonian Center for Astrophysics, 60 Garden Street, 
Cambridge, MA 02138}
 
\begin{abstract}

The standard spectral model for analyzing the soft component of
thermal emission from a thin accretion disk around a black hole is the
multi-temperature blackbody (MTB) model.  The widely used
implementation of this model, which is known as ``diskbb,'' assumes
nonzero torque at the inner edge of the accretion disk.  This
assumption is contrary to the classic and current literature on
thin-disk accretion, which advocates the use of a zero-torque boundary
condition.  Consequently, we have written code for a zero-torque
model, ``ezdiskbb,'' which we compare to the nonzero-torque model
diskbb by fitting {\it RXTE} spectra of three well-known black hole
binaries: 4U 1543-47, XTE J1550-564, and GRO J1655-40.  The chief
difference we find is that the zero-torque model gives a value for the
inner disk radius that is $\approx$ 2.2 times smaller than the value
given by diskbb.  This result has important implications, especially
for the determination of black-hole angular momentum and mass
accretion rate.

\end{abstract}


\section{Introduction}

One of the most widely used models in studies of the spectra of
accretion disks in black hole X-ray binaries is the multi-temperature
blackbody (MTB) model, which has been standard in the literature for
over 30 years \citep{pri72,sha73,pri81,fra92}.  In many systems, an
$L_{X} \propto T^{4}$ dependence has been observed (see, e.g.,
Gierli\'nski \& Done 2004), confirming a fundamental prediction of
such models and demonstrating that disks are an important and
prevalent accretion structure in black hole systems.  Fitting spectra
with the MTB model allows us to determine important properties of
accretion disks, including accretion rate, inner radius, and
temperature.  This information about the accretion disk can in
principle be used to deduce the angular momentum of the black hole
(e.g., Zhang, Cui, \& Chen 1997), and the radius at which an accretion
disk might be truncated in low accretion rate phases (e.g., Esin et
al.\ 2001).

Over the years, a number of different assumptions have been used in
deriving the MTB spectrum.  A standard assumption in the literature is
that a zero-torque boundary condition should be applied to the inner
edge of the accretion disk \citep{sha73,nov73}.  Gierli\'nski et
al. (1999) have shown, however, that the widely used MTB model known
as ``diskbb" (see, e.g., Mitsuda et al.\ 1984) that is found in
various X-ray spectral fitting packages assumes a nonzero torque at
the inner boundary of the disk.

In order to test the effects of zero-torque and nonzero-torque inner
disk boundary conditions, we have constructed disk models for each
scenario and fit a number of black hole binary spectra.  In section 2,
we outline the MTB model and illustrate the difference in the
temperature profiles that are obtained when the zero-torque inner
boundary condition is and is not used.  In section 3, we analyze the
spectra of three well-known black hole X-ray novae (BHXN): 4U1543-47,
XTE J1550-564, and GRO J1655-40.  Section 4 discusses the implications
of the results, and section 5 presents a summary and some conclusions.

\section{Multi-Temperature Disk Blackbody Model}

In this paper, we are interested in the multi-temperature blackbody
spectrum of a thin accretion disk around a compact star.  Following
standard treatments of this problem (e.g., Pringle 1981, Frank et al.
1992) we assume Newtonian gravity and take the disk to be in steady
state.  We limit ourselves to a brief discussion of the main results,
referring the reader to the literature cited above for details.

A thin accretion disk has a negligible radial pressure gradient.
Therefore, at each radius $R$, the gas orbits at essentially the
Keplerian angular frequency, $\Omega_K=(GM/R^3)^{1/2}$, where $M$ is
the mass of the central star.  Internal viscous torques in the
accreting gas transfer angular momentum outward, allowing the material
to spiral in.  These torques dissipate energy, and the dissipation
rate per unit surface area is given by
\begin{equation}
{D(R)=\frac{1}{2} \nu \Sigma (R \Omega^{\prime}_{K})^{2}} ,
\end{equation} 
where $\nu$ is the kinematic coefficient of viscosity, $\Sigma$ is the
surface density, $\Omega^{\prime}_{K} = d\Omega_K/dR=-3\Omega_K/2R$,
and the factor of $1/2$ takes into account the two disk surfaces, one
above and one below the mid-plane.  The mass accretion rate is given
by ${\dot{M} = 2 \pi R \Sigma (-v_R)} = {\rm constant}$, where $v_R$
is the radial drift velocity of the gas (negative for inward flow).
Conservation of angular momentum is then expressed by the condition
\begin{equation}
{-\nu \Sigma \Omega^{\prime}_K = -v_R \Sigma \Omega_K + \frac{C}{2\pi
R^3}} ,
\end{equation} 
where $C$ is a constant of integration.  To determine the value of
$C$, we apply a boundary condition on the disk; specifically, we
assume a value for the torque at the inner edge of the disk, located
at $R=R_{\rm in}$.  The standard choice is a zero-torque boundary
condition (e.g., Pringle 1981; Frank et al. 1992), which is
appropriate in the following two situations.

First, if the accretion disk is terminated on the inside by a star
spinning at a rate below the break-up limit, then there must be a
radius close to the surface of the star at which the angular velocity
of the orbiting gas reaches a maximum, meaning that $\Omega^{\prime}_K
= 0$.  At this radius, since the shear is zero, the viscous shear
stress must vanish.  Thus, if we identify this radius as the inner
edge of the disk $R_{\rm in}$, then the torque at the inner edge
clearly vanishes.  In this picture, the accretion disk is identified
as the region $R \geq R_{\rm in}$ and it has a zero-torque inner
boundary condition, while the region between $R_{\rm in}$ and the
surface of the star is called the boundary layer.  In what follows, we
ignore the boundary layer, whose radiation is typically much harder
than the disk emission.

The second situation in which the zero-torque boundary condition is
valid is when a thin accretion disk extends down to the marginally
stable orbit at radius $R_{\rm ms}$ around a black hole or very
compact neutron star.  In such a disk, gas spirals in slowly and
viscously down to $R=R_{\rm ms}$ and then free-falls rapidly with
negligible viscous interactions.  If the accretion disk is thin, the
transition from viscous to dynamical flow occurs quite suddenly
\citep{afs03}.  We may then identify $R_{\rm in}=R_{\rm ms}$ as the
inner edge of the disk, and we may safely assume a nzero-torque boundary
condition at this radius.

Since the torque at $R=R_{\rm in}$ vanishes, the right-hand side of
equation (2) is zero.  This determines the value of $C$, and it is
then straightforward to solve for $D(R)$.  Let us assume that the disk
is optically thick and that it radiates as a modified blackbody with a
radial temperature profile $T(R)$.  Then we have
\begin{equation}
\sigma \left[\frac{T(R)}{f}\right]^4 = D(R) =
{3GM\dot{M }\over8\pi R^3} \left[ 1-\left({R_{\rm in}\over R}
\right)^{1/2} \right],
\end{equation} 
where $\sigma$ is the Stefan-Boltzmann constant, and the spectral
hardening factor $f$ accounts approximately for the modification of
the optically thick disk emission from a pure blackbody.  The
modification could occur because of the combined effect of scattering and
absorptive opacity (e.g., Zavlin, Pavlov \& Shibanov 1996; Rutledge et
al. 1999; McClintock, Narayan \& Rybicki 2004), or due to
Comptonization \citep{shi95}.  Neither effect is fully described by a
simple rescaling of the temperature of the spectrum by a single factor
$f$, nor is $f$ expected to be independent of radius.  For simplicity,
however, we make both assumptions in writing equation (3).  Note that
for a canonical blackbody spectrum, $f=1$.

We refer to the temperature defined by the equation above as $T_{\rm
zt}$ to indicate that it is derived with the zero-torque boundary
condition.  Defining a dimensionless radius, $r = R/R_{\rm in}$, we
thus have
\begin{equation}
{T_{\rm zt}(r) = T_{\star} r^{-3/4} \left( 1 - r^{-1/2} \right)^{1/4}} ,
\qquad
{T_{\star} = f \left( \frac{3GM \dot{M}}{8\pi R^3_{in} \sigma} \right)
^{1/4}} .
\end{equation}  
The parameter $T_{\star}$ is, however, not a very convenient quantity
since the disk does not achieve this temperature at any radius.  Let
us, therefore, re-express equation (4) in terms of the maximum
temperature of the disk, $T_{\rm max}$:
\begin{equation}
{T_{\rm zt}(r) = 2.05 T_{\rm max} r^{-3/4} \left( 1 - r^{-1/2}
\right)^{1/4}}, \qquad T_{\rm
max} = 0.488 T_{\star} , \qquad r \equiv {R\over R_{\rm in}} .
\end{equation}  
Finally, by integrating the Planck function (with spectral hardening
factor $f$) over the area of the disk, and assuming a distance $D$ and
inclination $i$ for the source, we obtain the observed spectral flux
in units of erg s$^{-1}$ cm$^{-2}$ keV$^{-1}$:
\begin{equation}
{F_E = K \frac{4 \pi E^3}{h^3 c^2} \int_1^{\infty}{
\frac{r}{e^{E/kT_{\rm zt}(r)}-1} dr}} ,
\end{equation} 
where
\begin{equation}
{K = {1\over f^4}\left( \frac{R_{\rm in}}{D} \right)^2 \cos i} .
\end{equation}
From equations (5)\--(7), we see that the spectrum is determined by
one parameter, $T_{\rm max}$, and that it has a normalization
constant, $K$, which depends on four system-specific quantities: the
inner radius of the disk, the inclination of the disk, the distance to
the source, and the spectral hardening factor.

By integrating $D(R)$ over the entire disk (top and bottom surfaces)
from $R=R_{\rm in}$ to $R_{\rm out} \to \infty$, we may calculate the
total luminosity of the disk:
\begin{equation}
L_{\rm disk} = {GM\dot M\over 2R_{\rm in}} =
73.9 \sigma \left( \frac{T_{\rm max}}{f} \right)^4 R_{\rm in}^2
\qquad {\rm (zero~torque)}.
\end{equation}
Note that only half the gravitational binding energy of the accreting
gas is radiated by the disk; the other half survives as kinetic energy
of the gas at the inner edge of the disk.  In the case of accretion
onto a star, some or all of this residual energy is dissipated in the
boundary layer as the gas spins down to merge with the surface of the
star (see Kluzniak 1987; Popham \& Narayan 1995 for an estimate of the
fraction of this energy that is radiated for a given stellar spin).
In the case of accretion onto a black hole, the energy remains in the
gas and falls into the hole.

The above discussion is for a disk with a zero-torque inner boundary
condition.  In some cases, however, a disk may have a finite torque at
its inner edge.  For instance, Popham \& Narayan (1991) and
Paczy\'nski (1991) showed that a star that has been spun up to breakup
can lose angular momentum to a surrounding accretion disk via a torque
at the interface.  The torque can (in principle) have any magnitude.
In the case of a disk around a black hole, some authors have recently
discussed the possibility that magnetic fields may induce a nonzero
torque at the marginally stable orbit \citep{gam99,ago00,rey01}.
Afshordi \& Paczy\'nski (2003) argue that the magnitude of the torque
at $R_{\rm ms}$ depends on the vertical thickness of the disk: whereas
a thick disk could have a substantial torque, a thin disk should at
most have only a very weak torque (see also Armitage, Reynolds \&
Chiang 2001).

The study of disks with nonzero torque is hampered by the fact that
there is no characteristic or ``natural'' magnitude that one can
select for the torque.  A complete analysis will thus need to include
the strength of the torque as a second parameter.  In this paper, we
consider a particular choice for the torque, corresponding to the case
when the integration constant $C$ in equation (2) is zero.  This is to
be viewed as just an example, though as we discuss in \S 3.1, this
particular case is historically important.  When $C=0$, the
temperature profile takes the simple form
\begin{equation}
        T_{\rm st}(r) = T_{\star} r^{-3/4} = T_{\rm max} r^{-3/4} ,
\end{equation} 
where $T_{\star}$ is again defined as in equation (4), and the
subscript in $T_{\rm st}$ stands for ``standard torque."  The profiles
$T_{\rm zt}(r)$ and $T_{\rm st}(r)$ are compared in Figure 1.  We see
that $T_{\rm zt}(r)$ goes to zero at the inner edge of the disk since
the torque vanishes there, whereas $T_{\rm st}(r)$ reaches its maximum
value at the inner edge because the torque is maximum.

The spectrum of a disk with the standard-torque temperature profile is
given by equation (6), except that $T_{\rm zt}(r)$ is replaced by
$T_{\rm st}(r)$.  The total disk luminosity is
\begin{equation}
 L_{\rm disk} = {3GM\dot M\over 2R_{\rm in}} = 12.6 \sigma
\left(\frac{T_{\rm max}}{f} \right)^4 R_{\rm in}^2 \qquad {\rm
(standard~torque)}.
\end{equation}
Note that for given $M$, $\dot{M}$, and $R_{\rm in}$, this disk model
has approximately twice the maximum temperature and exactly three
times the luminosity of a disk with zero torque.  The extra luminosity
is because of the work done by the nonzero torque at the inner edge.
Figure 2 shows the MTB spectra of the two temperature models presented
in Figure 1.  There is a significant difference between the two
spectra, especially for photon energies greater than 1 keV (for this
particular example), where the $T_{\rm st}$ model has substantially
more emission.

\begin{figure}
\plotone{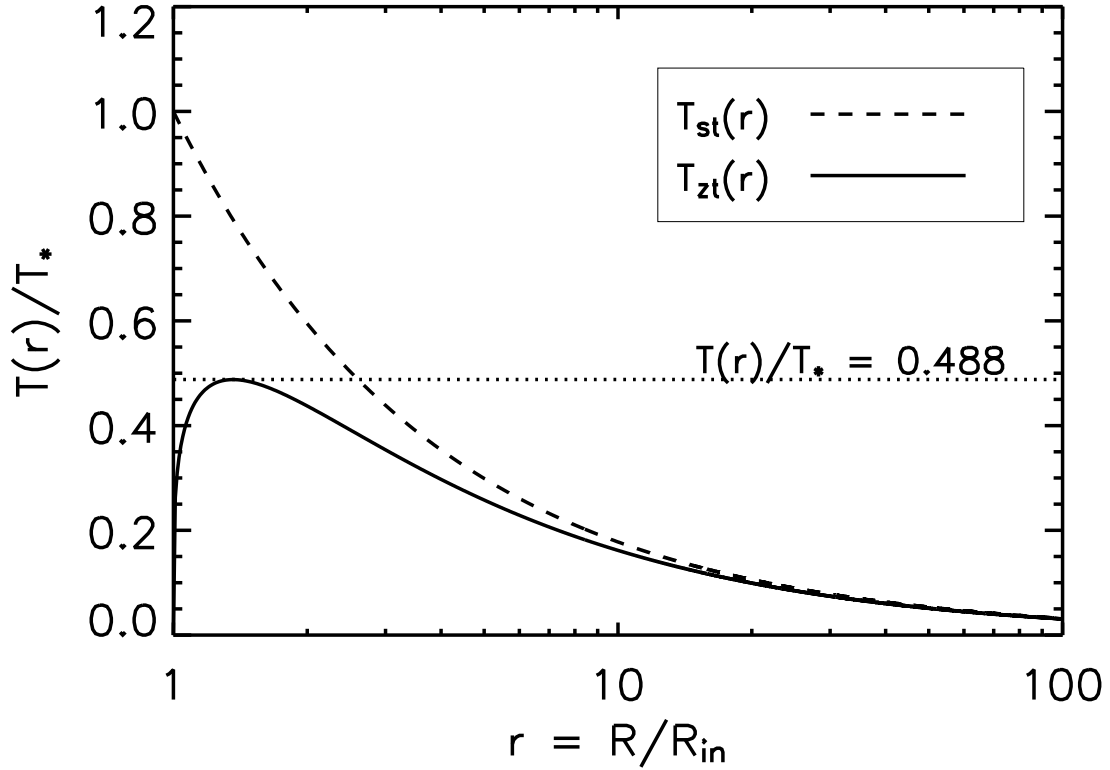}
\caption{Disk temperature as a function of radius for the two models
discussed in \S 2.  The vertical axis is in units of $T/T_{\star}$,
while the horizontal axis is in units of $r=R/R_{\rm in}$.  The solid
curve shows the profile $T_{\rm zt}(r)$ of a disk with a zero-torque
boundary condition at the inner edge of the disk (eq. 4).  The
dashed curve shows the profile $T_{\rm st}(r)$ of the standard-torque
model (eq. 9).}
\end{figure}

\begin{figure}
\plotone{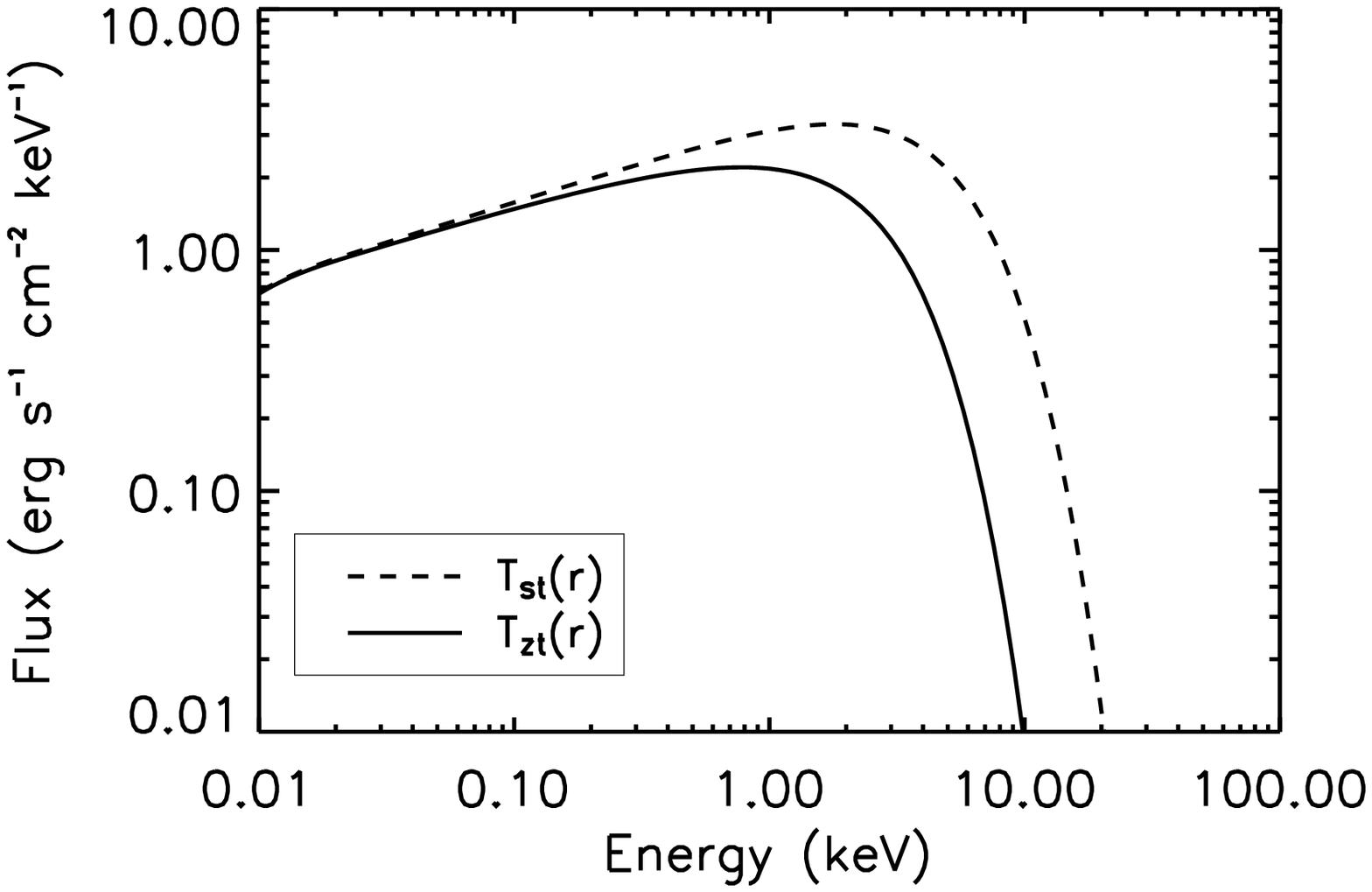}
\caption{MTB disk spectra corresponding to the two temperature
profiles shown in Figure~1.  The following parameters have been
assumed: $T_{\star} = 2$ keV, $R_{\rm in}$ = 10 km, $D$ = 10 kpc, $i$
= 0 degrees, $f=1$.  Note the large difference between the two
spectra, especially for photon energies above 1 keV.}
\end{figure}

\section{Analysis and Results}

\subsection{Disk Models}

The X-ray spectral fitting package XSPEC (Arnaud 1996) contains
diskbb, a widely used version of the MTB spectral model.  As we stated
in the introduction, Gierli\'nski et al. (1999) have pointed out that
diskbb assumes a nonzero torque at the inner boundary of the accretion
disk.  The help manual on diskbb within XSPEC and the papers to which
it refers (Mitsuda et al. 1984; Makishima et al. 1986) are ambiguous
as to whether Gierli\'nski et al.'s assertion is true.  The latter
papers refer to Pringle (1981), which describes both the zero-torque
temperature profile (eq. 4 in our paper), and the standard-torque
profile (eq. 9).  In private correspondence, Makishima, one of the
originators of the diskbb code, confirmed that the standard-torque
condition is assumed by diskbb.  We decided to check this for
ourselves.  We wrote a code called diskbbcheck, which explicitly uses
$T_{\rm st}(r)$ as the temperature profile and, as described below, we
compared the results obtained using diskbbcheck against those obtained
with diskbb.

We also wrote code for another model, which we call ezdiskbb.  This
model corresponds to the zero-torque boundary condition at the inner
edge of the disk, and makes use of the temperature profile $T_{\rm
zt}(r)$.  We used this code to compare the results of spectral fits
done with and without the zero-torque inner boundary condition, as
discussed below.

Each of the models, diskbb, diskbbcheck, and ezdiskbb has two
adjustable parameters.  The first is the maximum temperature in the
disk, which we call $T_{\rm max}$.  When the standard-torque boundary
condition is assumed, $T_{\rm max}$ is identical to $T_{\star}$, the
temperature at the inner edge of the accretion disk (Fig. 1).  The
second parameter in all three models is a normalization constant, $K$,
which is defined as follows:

\begin{equation}{K= \frac{1}{f^4} \left( \frac{R_{\rm in}/1 \mbox{km}}
	{D/10 \mbox{kpc}} \right)^2 \cos{i}}  .
\end{equation}  
The current version of diskbb implicitly assumes a spectral hardening
factor of $f=1$ in its normalization.  A summary of the three models
is given in Table 1.

\begin{deluxetable}{cccc}
\tablecolumns{4}
\tablewidth{0pt}
\tablecaption{MTB Models}
\tabletypesize{\footnotesize}
\tablehead{
	\colhead{Model} &
	\colhead{BC} &
	\colhead{Description} &
	\colhead{References}	
	}

\startdata

diskbb	& Standard-torque & Standard XSPEC model & Mitsuda et al. (1984); 
	Makishima et al. (1986) \\

diskbbcheck & Standard-torque & Confirmed BC in diskbb	& This work	\\

ezdiskbb & Zero-torque	& Alternative to diskbb  & This work	\\

\enddata

\end{deluxetable}

\subsection{Analysis}

We compared the performance of the two models by analyzing data for
three BHXN that were obtained using the Proportional Counter Array
(PCA) aboard the {\it Rossi X-ray Timing Explorer} (RXTE).  The PCA
consists of five xenon-filled proportional counter units (PCUs) that
have a total effective area of 6200 cm$^2$ at 5 keV \citep{jah96}.
The PCA is sensitive over 2-60 keV, with an effective energy range of
$\sim$ 2.5-20 keV and an energy resolution of $\sim$ 17\% at 5 keV
\citep{sob00a}.  We also used lightcurves obtained using RXTE's
All-Sky Monitor (ASM), which scans $\sim$ 80\% of the sky during every
orbit \citep{sob00a}.  The ASM has three energy channels: 1.5-3 keV,
3-5 keV, and 5-12 keV and is very useful for identifying and
monitoring transient sources like BHXN.

In our analysis, we used PCA data for three BHXN: 4U1543-47 (hereafter
U1543), XTE J1550-564 (hereafter J1550), and GRO J1655-40 (hereafter
J1655).  The BH mass, distance, and inclination have been determined
through optical observations for each of these objects.  These
properties are given in Table 2.

\begin{deluxetable}{cccc}
\tablecolumns{4}
\tablewidth{0pt}
\tablecaption{BHXN Properties}
\tablehead{
\colhead{} & 
\colhead{$M_{BH}$ ($M_{\odot}$)} & 
\colhead{$D$ (kpc)} & 
\colhead{$i$ ($\degr$)}
}
\startdata
\\
73.5$^{+1.9}_{-2.7}$$^b$  \\
0.08$^c$ \\
4U 1543-47 & 9.4 $\pm$ 2.0\tablenotemark{a} & 7.5 $\pm$ 
1.0\tablenotemark{a} & 20.7 $\pm$
	1.0\tablenotemark{a} \\
XTE J1550-564 & 10.56$^{+1.02}_{-0.88}$\tablenotemark{b} & 
5.9$^{+1.7}_{-3.1}$\tablenotemark{b} &
	73.5$^{+1.9}_{-2.7}$\tablenotemark{b}  \\
GRO 1655-40 & 7.02 $\pm$ 0.22\tablenotemark{c} & 3.2 $\pm$ 
0.2\tablenotemark{d} & 69.50 $\pm$
	0.08\tablenotemark{c} \\
\enddata
\tablenotetext{a}{Orosz et al. (1998); Orosz (2003)}
\tablenotetext{b}{Orosz et al. (2002)}
\tablenotetext{c}{Orosz \& Bailyn (1997)}
\tablenotetext{d}{Hjellming \& Rupen (1995)}
\end{deluxetable}

These three objects have all had recent outbursts that have been
analyzed in detail using RXTE observations.  Park et al. (2003)
studied the 2002 outburst of U1543; Sobczak et al. (2000) analyzed the
outburst of J1550 in 1998-1999; and Sobczak et al. (1999) analyzed
J1655's 1996-1997 outburst.  The ASM lightcurves of each of these
outbursts, along with their hardness ratios (HR2) versus time, are
shown in Figure 3.  The values for HR2 correspond to the 5-12 keV
intensity divided by the 3-5 keV intensity, as measured by the ASM.
For each of these sources, we chose 10 observations to analyze.  All
of these observations were chosen when the source was in what is
canonically known as its ``high/soft" state, during which the soft
thermal component from a hot disk dominates the spectrum
\citep{mcc04}.  This state corresponds approximately to a hardness
ratio of 0.5.  The times of our chosen observations are denoted by the
dashed vertical lines in Figure 3.

\begin{figure}
\epsscale{0.45}
\plotone{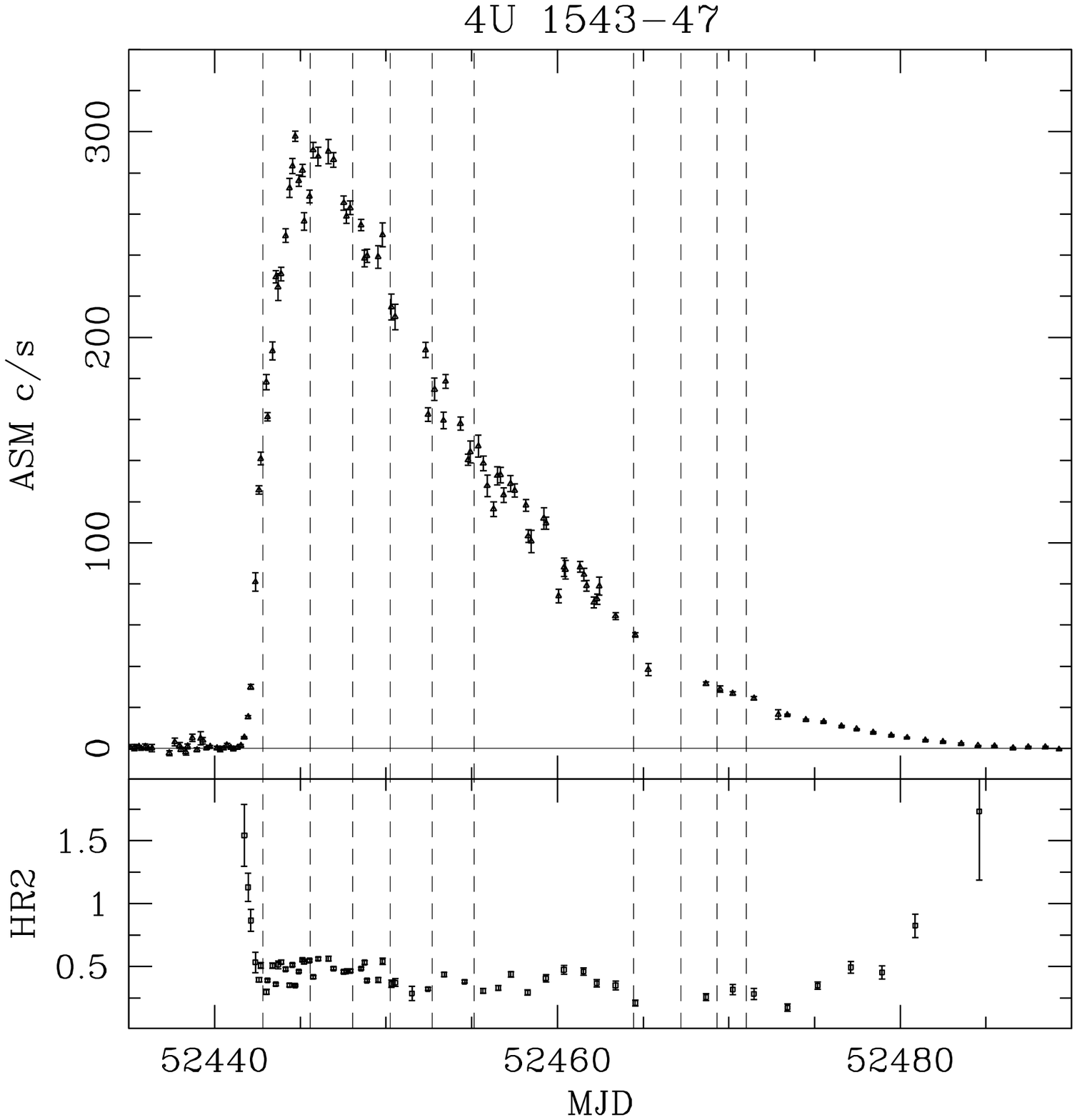}
\plotone{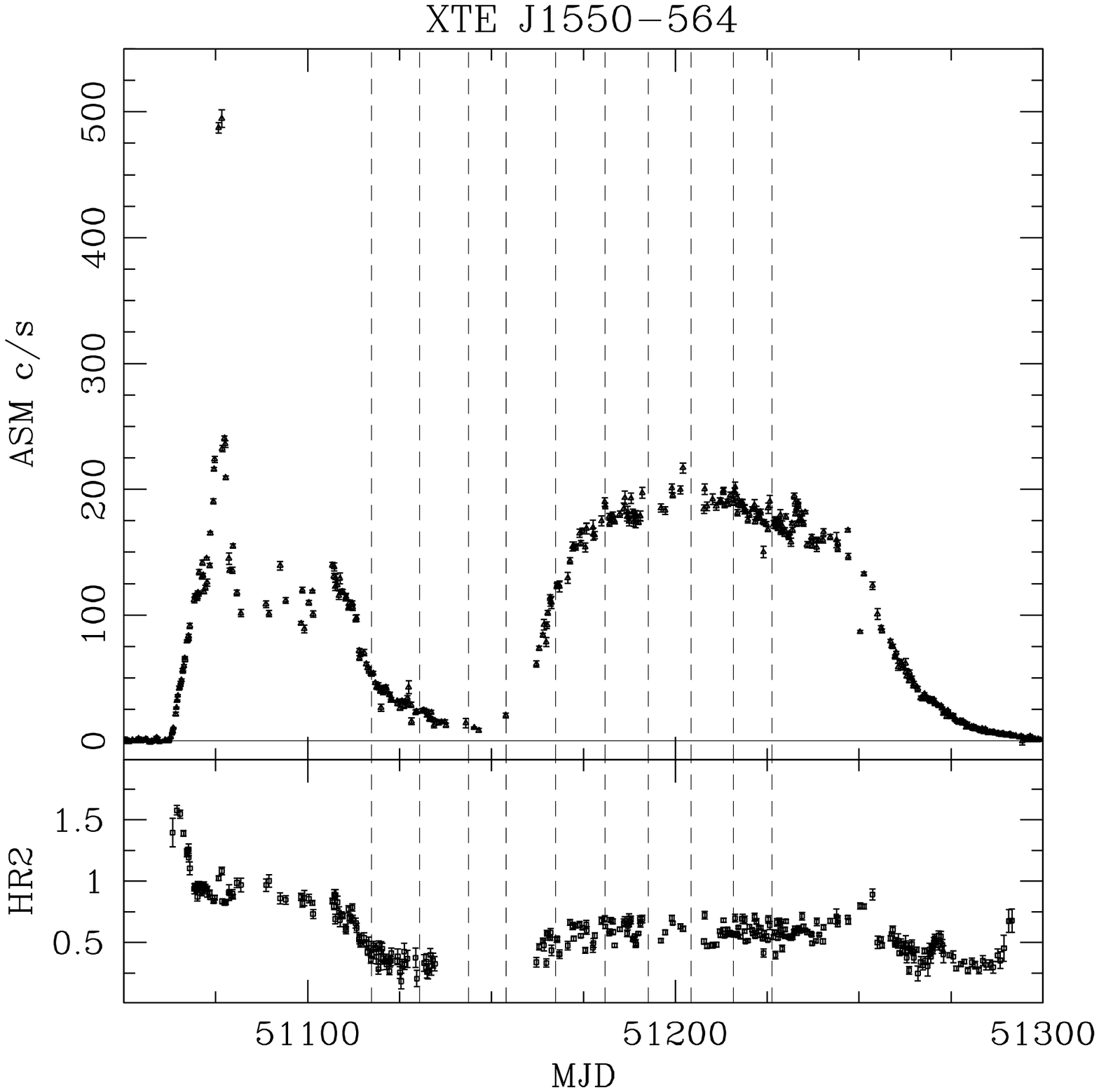}
\plotone{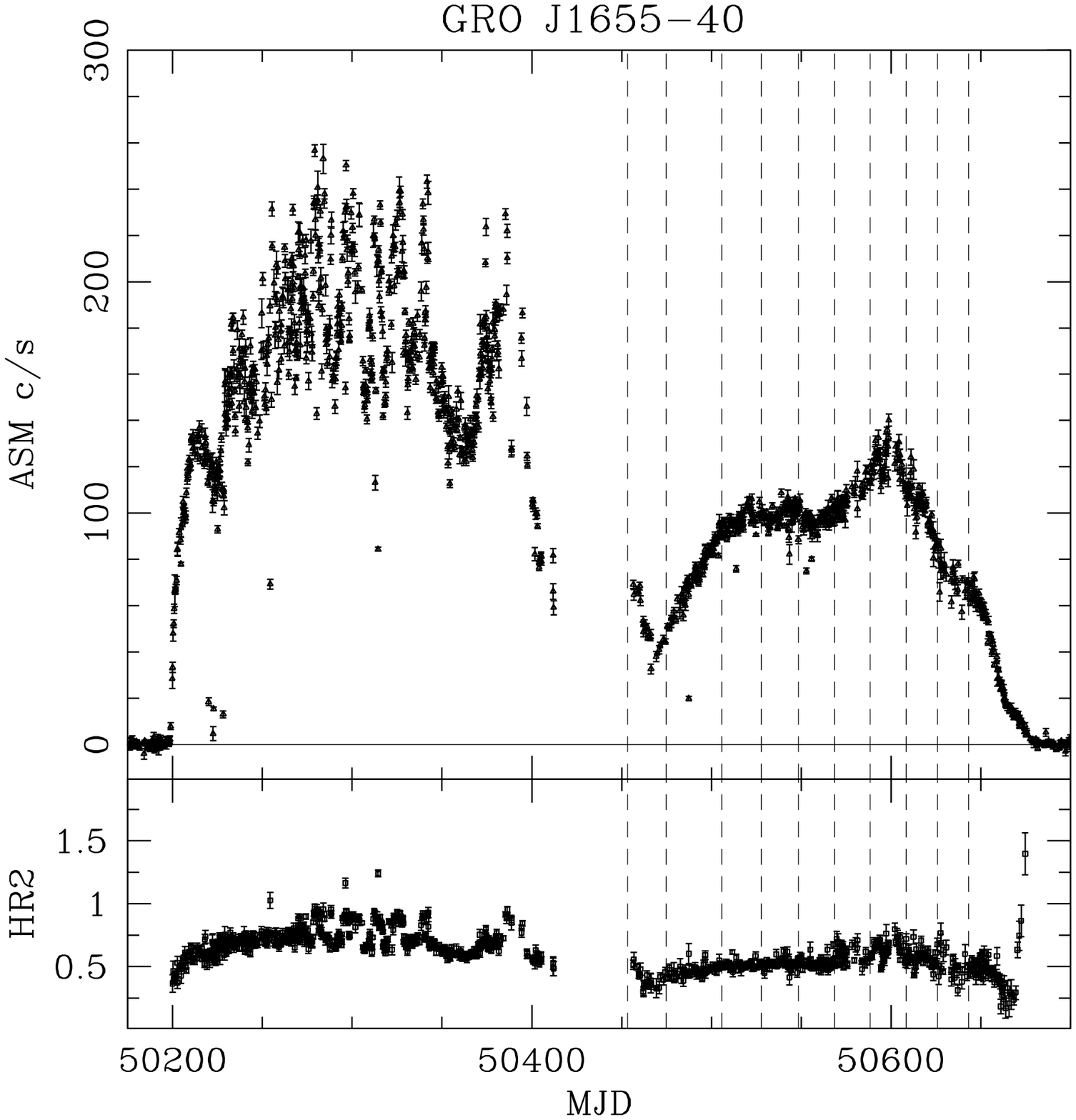}
\caption{The 1.5-12 keV intensities and hardness ratios of 4U 1543-47,
XTE J1550-564, and GRO J1655-40 plotted versus time, as observed by
the RXTE All-Sky Monitor.  An X-ray intensity of 1 Crab corresponds to
75.5 ASM c/s.  The hardness ratio is the ratio of the 5-12 keV
intensity to the 3-5 keV intensity. The time axis is in units of MJD =
JD $-$ 2,400,000.5.  The dashed lines in each plot correspond to the
10 observation times that were analyzed in this paper for each source.
Note that all the selected observations occurred when the sources were
in their high/soft states, characterized by a hardness ratio of
approximately 0.5.}
\end{figure}

In fitting the spectra of these sources, we followed as closely as
possible the procedures described in the papers cited above.  In
addition to the MTB models, we used four other XSPEC models in our
fits.  Primary among these was the power-law model, which has two
parameters: the photon index, $\Gamma$, and a normalization constant,
which we will call $K_{\rm PL}$.  The next component that we used was
an interstellar absorption model, which has one parameter: the
hydrogen column density, $N_{\rm H}$, in units of atoms cm$^{-2}$.
The third component in our fits was a smeared absorption edge model
(``smedge"), which has three parameters: the threshold energy in keV,
the smearing width in keV, and the maximum absorption factor at
threshold, $\tau_{\rm FE}$ (the smedge model does have one additional
parameter, the index for photoelectric cross-section, but this
parameter always remains fixed at a value of $-2.67$).  Our final
spectral component was a Gaussian, which was used to fit an Fe line
feature for U1543 and J1550, but not J1655.  The three parameters in
this model are the line energy in keV, the line width in keV, and the
normalization, which we will call $K_{\rm line}$.

Because our focus was on the MTB models, we froze a number of the
extraneous parameters in these other spectral components to minimize
their influence on our fits.  We fixed the parameters at average
values obtained from the previously cited papers.  Whenever a
parameter was frozen, it was frozen at the same value for all 10
spectra for each source.  Table 3 shows which parameters were allowed
to float and which were frozen for each object, along with the values
at which they were frozen.  A blank field in the table indicates that
the given spectral component was not used in fits for that source
(i.e., the Gaussian for J1655).  Note that the disk models and the
power-law model are not included in the table, because the parameters
from both of these models were always allowed to float.

We analyzed PCA data from all available PCUs; this was generally 3 of
the PCUs for each source.  Prior to doing our fits, we added 1\%
systematic errors to all of the energy channels of our spectra.  We
obtained the response matrices for our fits from the HEASARC archives.
Table 3 also gives the energy range over which the spectra were fit
for each source.  Again, the guidelines in the cited papers were
followed.  For J1550 and J1655, Sobczak et al. (1999, 2000) adopted an
energy range of 2.5-20 keV for their fits, which was appropriate for
the earlier PCA response matrices that they used.  However, we were
unable to obtain acceptable fits over this energy range using the
current response matrices.  We therefore raised the lower bound on the
energy range to 3.0 and 2.8 keV for J1550 and J1655, respectively.

\begin{deluxetable}{ccccccccc}
\tablecolumns{9}
\tablewidth{0pt}
\tabletypesize{\footnotesize}
\tablecaption{Spectral Fit Components\tablenotemark{a}}

\tablehead{
\multicolumn{1}{c}{} &
\multicolumn{1}{c}{Absorption} &
\multicolumn{3}{c}{Smedge} &
\multicolumn{3}{c}{Gaussian} &
\multicolumn{1}{c}{}}

\startdata
 & $N_{H}$ & Energy & Width & $\tau_{FE}$ & Energy & Width &
 	$K_{line}$ & Range of Fits \\
 & (10$^{22}$ cm$^{-2}$) & (keV) & (keV) & & (keV) & (keV) & & (keV) \\

 \hline

4U 1543-47\tablenotemark{b} & 0.40 & 7.4 & 7.0 & Float & 6.5 & 1.2 & 
Float & 2.9-25 \\
XTE J1550-564\tablenotemark{c} & 2.0 & 8.6 & 7.0 & Float & 6.5 & 1.2 & 
Float & 3.0-20  \\
GRO 1655-40\tablenotemark{d} & 0.89 & 8.0 & 7.0 & Float & & & & 2.8-20 \\
\enddata

\tablenotetext{a}{All values indicate frozen values for that parameter 
for all fits.  ``Float" indicates a
fit parameter that was always allowed to vary.  The blank cells for 
J1655 indicate that a Gaussian was not
used in the fits for that object.}
\tablenotetext{b}{Values used for fits were obtained from Park et al. 
(2003)}
\tablenotetext{c}{Values used for fits were obtained from Sobczak et al. 
(2000)}
\tablenotetext{d}{Values used for fits were obtained from Sobczak et al. 
(1999)}

\end{deluxetable}

\subsection{Results}

We first determined the boundary condition used in the standard XSPEC
model, diskbb, by fitting our 10 spectra of U1543 with both diskbb and
diskbbcheck.  For all 10 spectra, we obtained identical results with
the two models.  We therefore confirmed that diskbb is derived using
the standard-torque boundary condition.

We next explored the effects of using the zero-torque boundary condition
by fitting the spectra of all three sources using ezdiskbb and diskbb
and comparing the fit parameters.  We found that ezdiskbb gave values
for $T_{\rm max}$ that were $\approx$ 5\% lower than those given by
diskbb when averaged over all of our fits.  Figure~4 illustrates this
effect, and Table~4 gives the average ratios between the $T_{\rm max}$
values obtained using diskbb and ezdiskbb for each of our sources, as
well as for the extreme range of the ratios between the two models.  In
Figure 4, we have plotted the values of $T_{\max}$ for a spectral
hardening factor of $f=1.7$, the value that is quoted in Shimura \&
Takahara (1995) and that is generally accepted throughout the literature
(see the discussion in \S~4 below).  The hardening factor relates the
observed temperature $T_{\rm max}$ to the effective temperature: $T_{\rm
eff} = T_{\rm max}/f$.  Note that the hardening factor does not affect
the {\it ratio} between the diskbb temperature and the ezdiskbb
temperature, as long as the same value of $f$ is used in both.

\begin{figure}
\epsscale{0.45}
\plotone{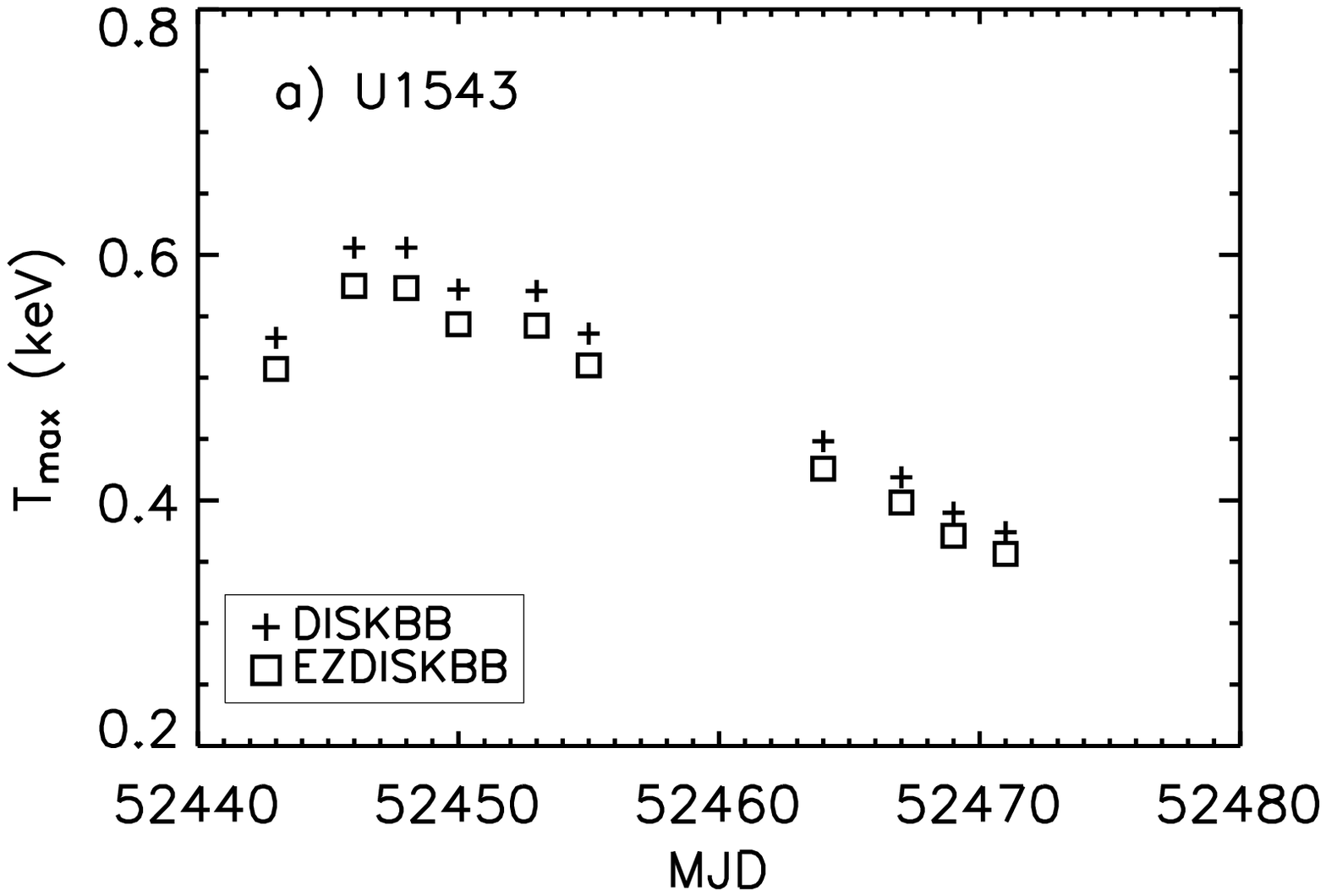}
\plotone{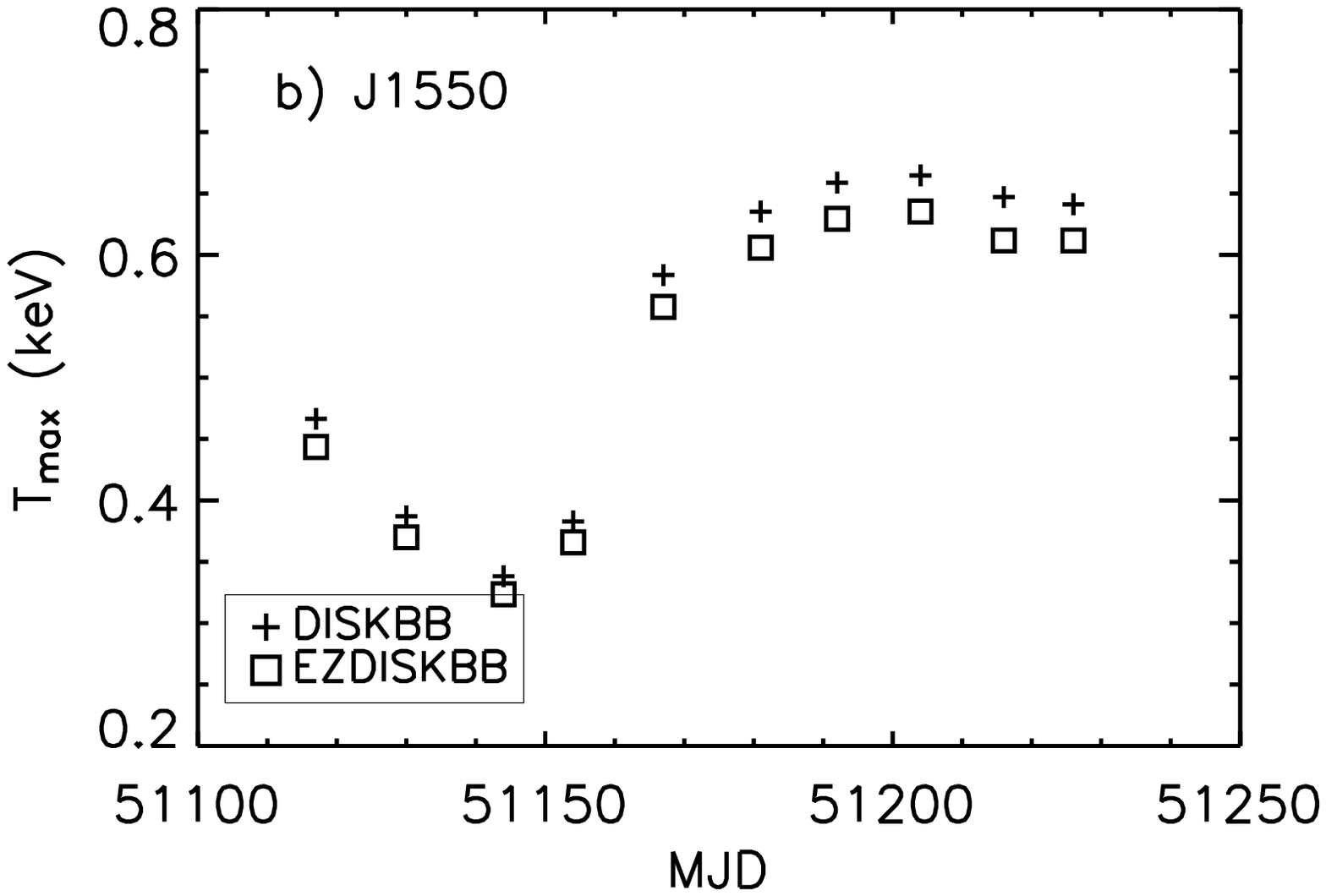}
\plotone{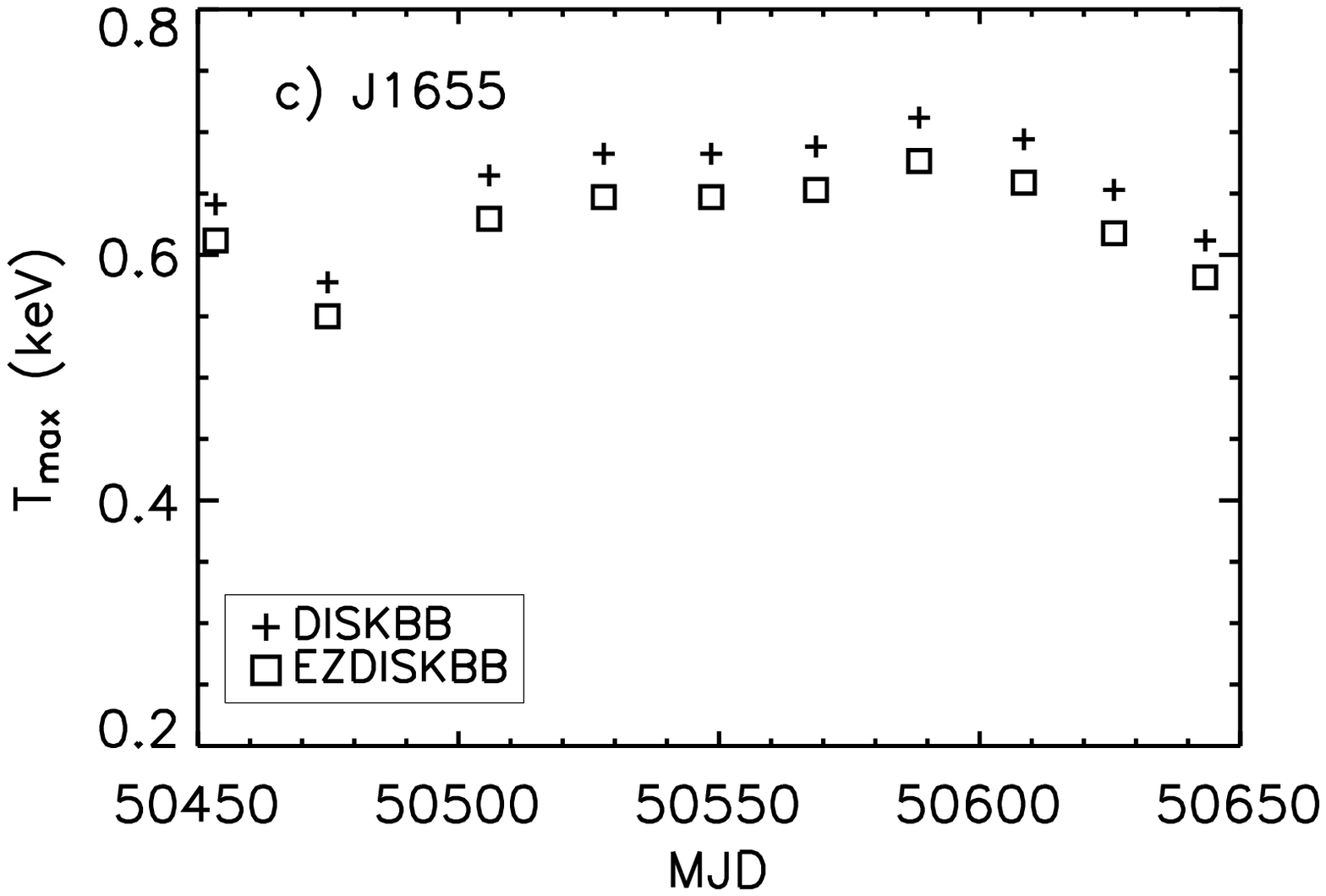}
\caption{\small{$T_{\rm max}$ versus time for the 10 spectra
from each of our BHXN, assuming a spectral hardening factor of
$f=1.7$.  The horizontal axis gives the date of each observation in
units of MJD = JD $-$ 2,400,000.5.  The plus symbols indicate the
values of $T_{\rm max}$ obtained from fits using diskbb, and the
squares indicate $T_{\rm max}$ values from fits using ezdiskbb.  We
see that ezdiskbb, which assumes a zero-torque boundary condition at the
inner edge of the accretion disk, gives values for the maximum
temperature in the disk that are always about 5\% lower than those
given by diskbb, which assumes the standard-torque inner boundary
condition.  This ratio is independent of $f$.}}
\end{figure}

A much more significant effect of using the zero-torque condition was
that the values of the inner radius of the disk, $R_{\rm in}$, were
reduced by more than a factor of 2.  Specifically, ezdiskbb gave
values for $R_{\rm in}$ that were a factor of 2.17 smaller on average
than those given by diskbb, with the smallest difference between the
two models being a factor of 2.15.  This result is shown in Figure 5.
The average ratios between the values of $R_{\rm in}$ for the two
models for each source are given in Table 4.  The ratios are
independent of $f$.

This large difference in the values of $R_{\rm in}$ between the two
models can be explained as follows.  In fitting the disk spectrum, the
form of which is given in equation (6), XSPEC must fit the shape of the
model spectrum --- as determined for instance by the peak of the
emission --- to the observed spectrum.  The shape of the MTB model
spectrum depends only on the maximum temperature in the disk, and we saw
above that diskbb reproduces the observational spectrum with a $T_{\rm
max}$ that is about 1.05 times larger than the corresponding $T_{\rm
max}$ for ezdiskbb.  XSPEC must also fit the overall normalization of
the model spectrum to the observed one in order to achieve the same
luminosity.  Equating the luminosities for the two model spectra using
eqs. (8) and (10) gives:

\begin{equation} \left[ 73.9 \sigma \left( \frac{T_{\rm max}}{f} 
\right)^4 R_{\rm in}^2 \right]_{\rm zt} =
		\left[ 12.6 \sigma \left( \frac{T_{\rm max}}{f}
		\right)^4 R_{\rm in}^2 \right]_{\rm st} .
\end{equation}  
The right-hand side of equation (12) corresponds to diskbb with its 1.05
times larger value of $T_{\rm max}$.  In order to maintain the equality,
this temperature difference requires that $R_{\rm in}$ be a factor of
$\sim2.2$ larger for diskbb than for ezdiskbb, in agreement with the
ratio obtained from the fits (Table~4).

\begin{figure}
\epsscale{0.45}
\plotone{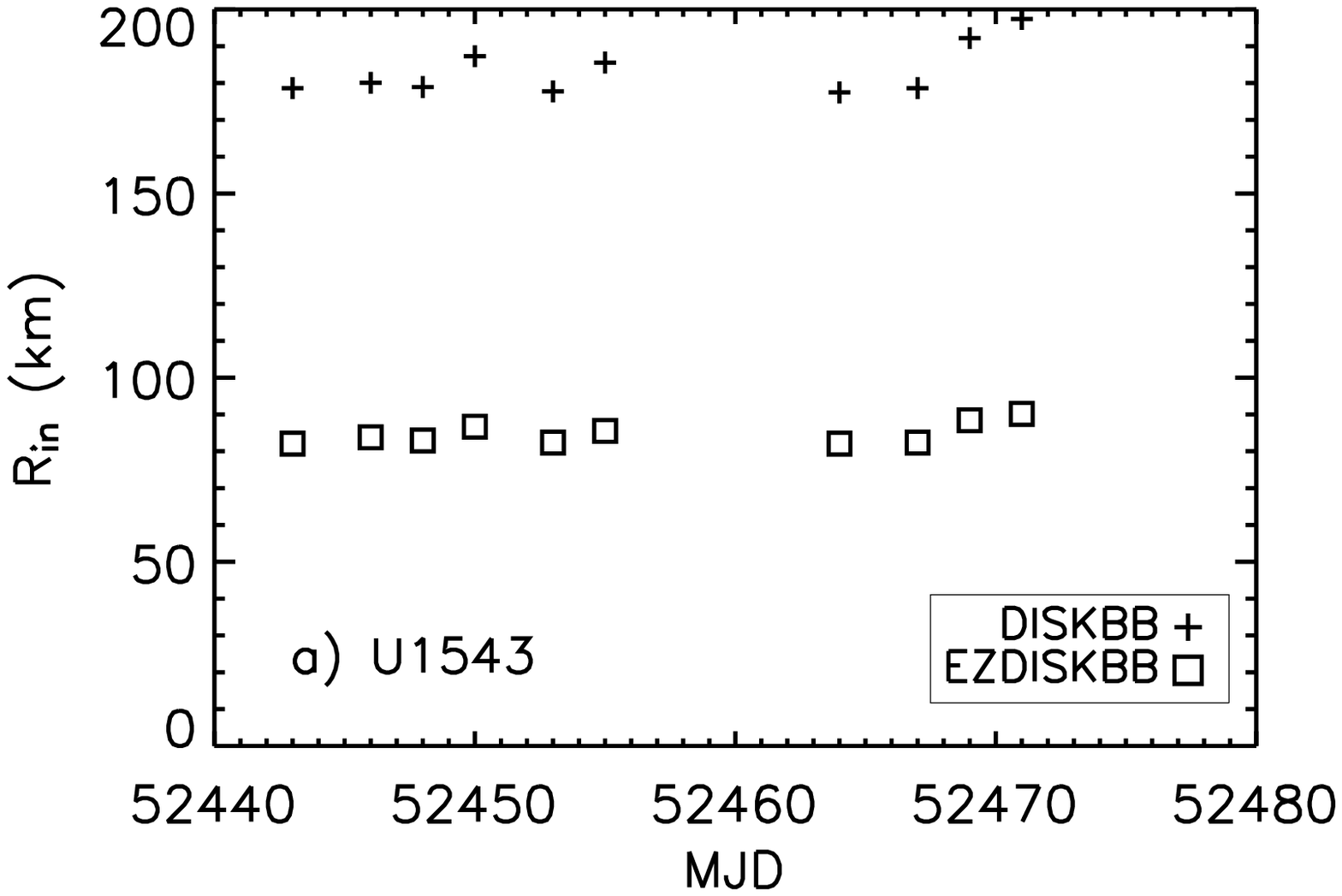}
\plotone{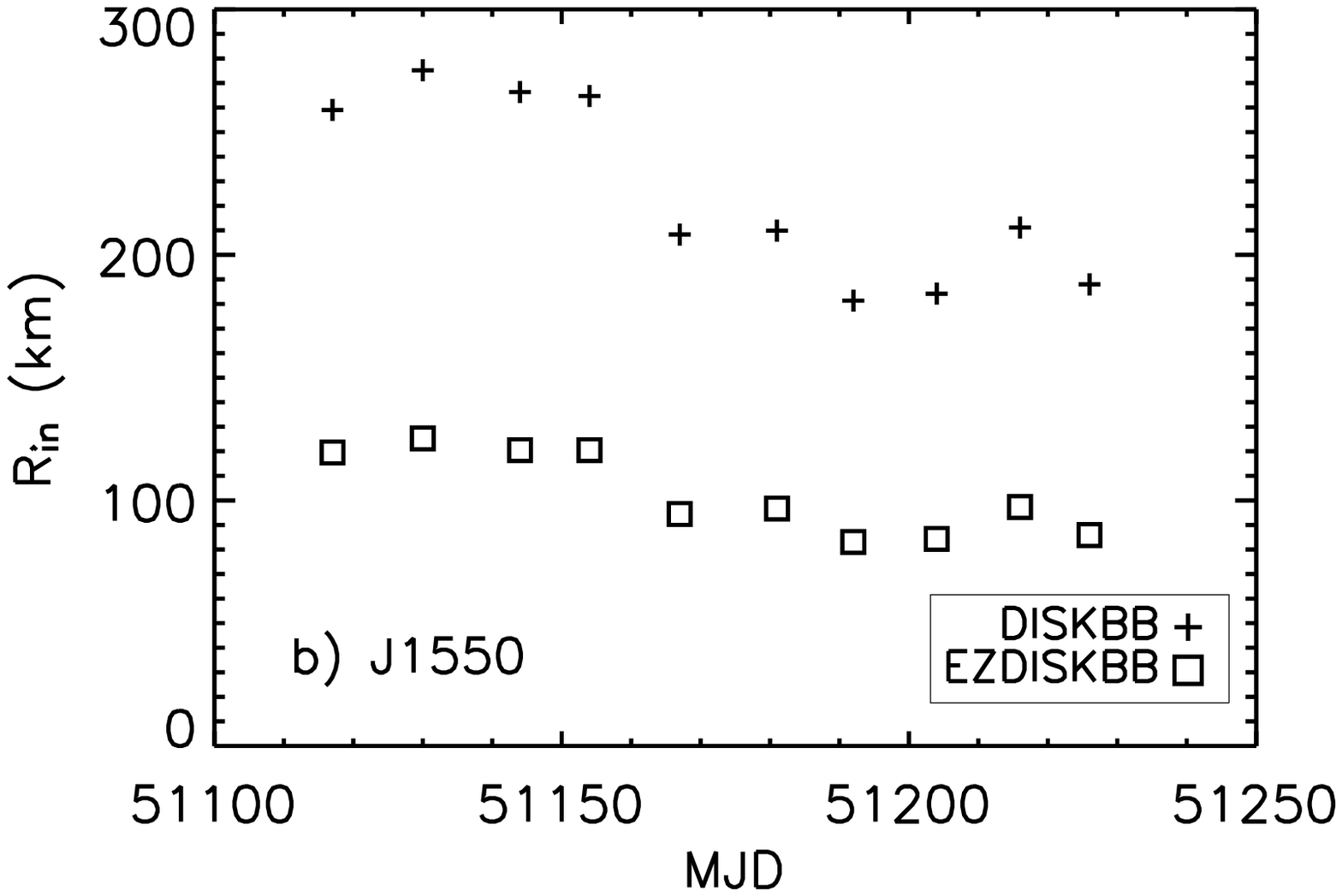}
\plotone{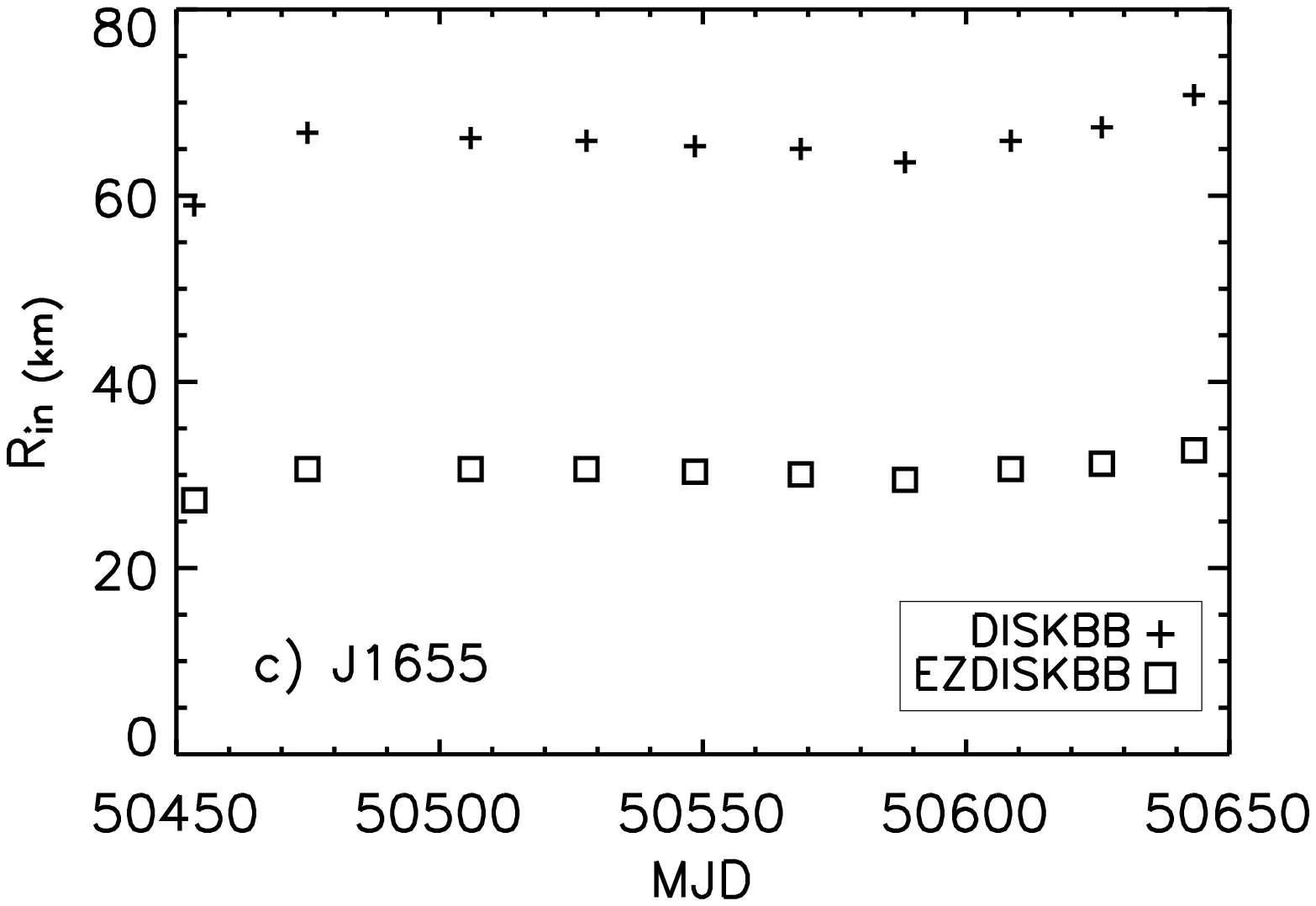}
\caption{\small{$R_{\rm in}$ versus time for the 10 spectra from each
of our BHXN, assuming a spectral hardening factor of $f=1.7$.  The
plus symbols indicate the values of $R_{\rm in}$ obtained from fits
using diskbb, and the squares indicate $R_{\rm in}$ values from fits
using ezdiskbb.  We see that ezdiskbb gives values for the inner
radius of the disk that are a factor of $\approx$ 2.2 smaller than
those given by diskbb; this ratio is independent of $f$.  Therefore,
whether one uses the zero-torque or the standard-torque boundary
condition has a large effect on the determination of the inner radius
of the accretion disk.}}
\end{figure}

We can also estimate the effect on the estimated mass accretion rate,
$\dot{M}$.  Comparing equations (8) and (10), we see that for a given
$M$, $\dot M$ and $R_{\rm in}$, the luminosity of the zero-torque
model is 3 times less than that of the standard-torque model.
Further, we have just shown that $R_{\rm in}$ for the former model is
2.2 times less compared to the latter model.  Therefore, in order to
have the same luminosity, the value of $\dot M$ with the
standard-torque model (diskbb) should be $2.2/3 \sim 0.7$ times the
value of $\dot M$ with the zero-torque model (ezdiskbb).  The results
of the fits to data are shown in Figure 6, and the average differences
in $\dot{M}$ between diskbb and ezdiskbb are given in Table~4.  The
results are consistent with the theoretical expectation.

\begin{figure}
\epsscale{0.45} \plotone{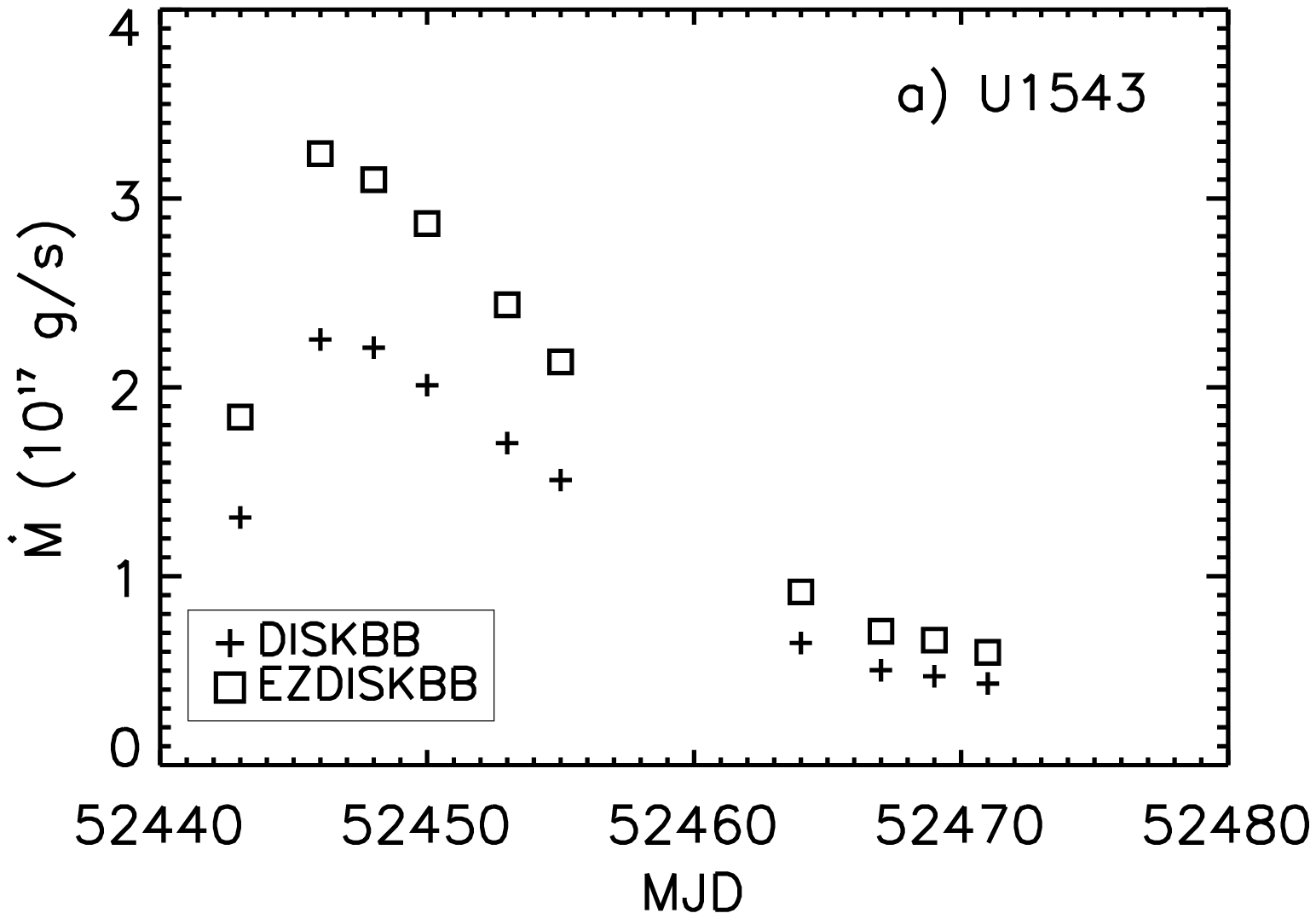} \plotone{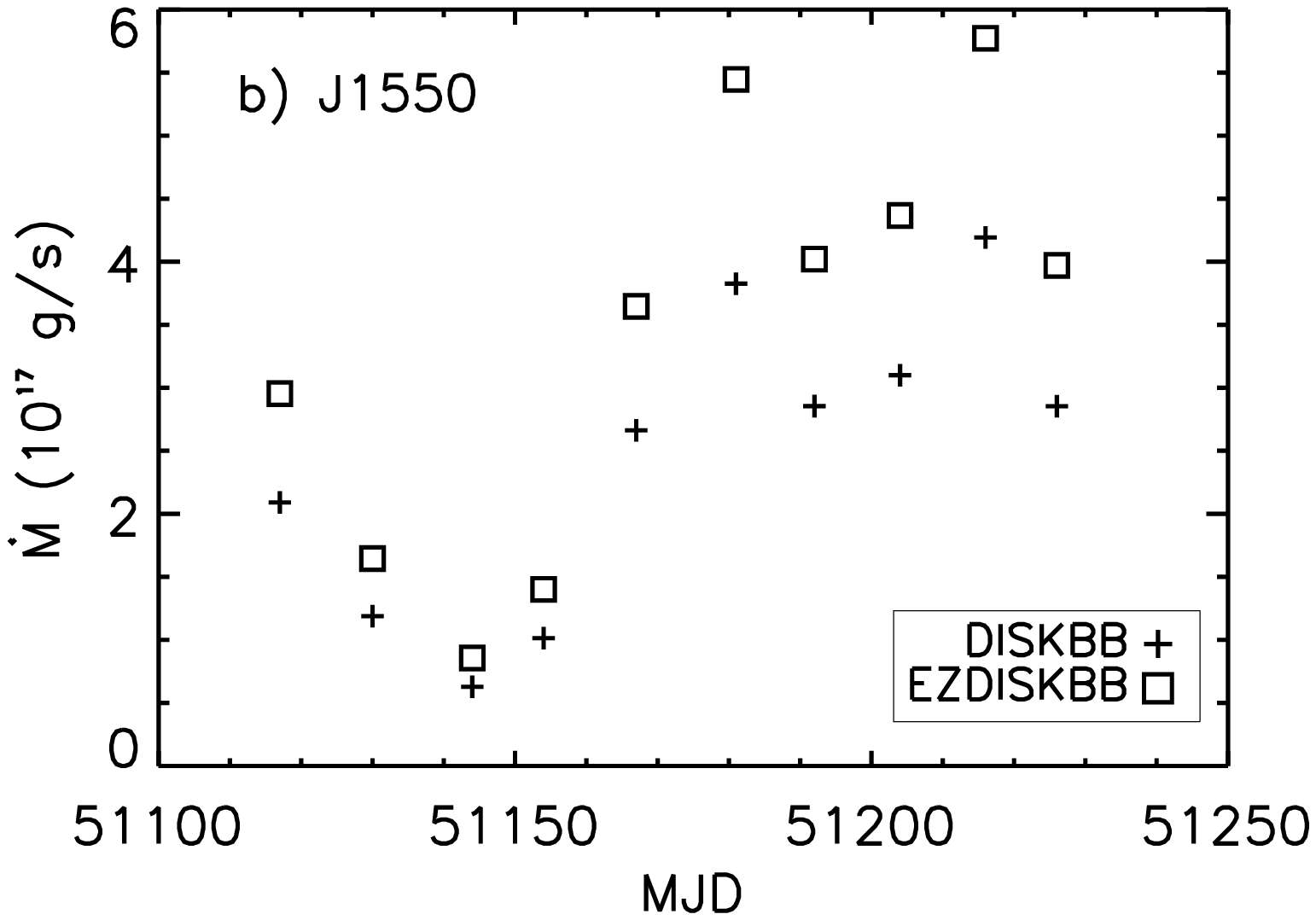} \plotone{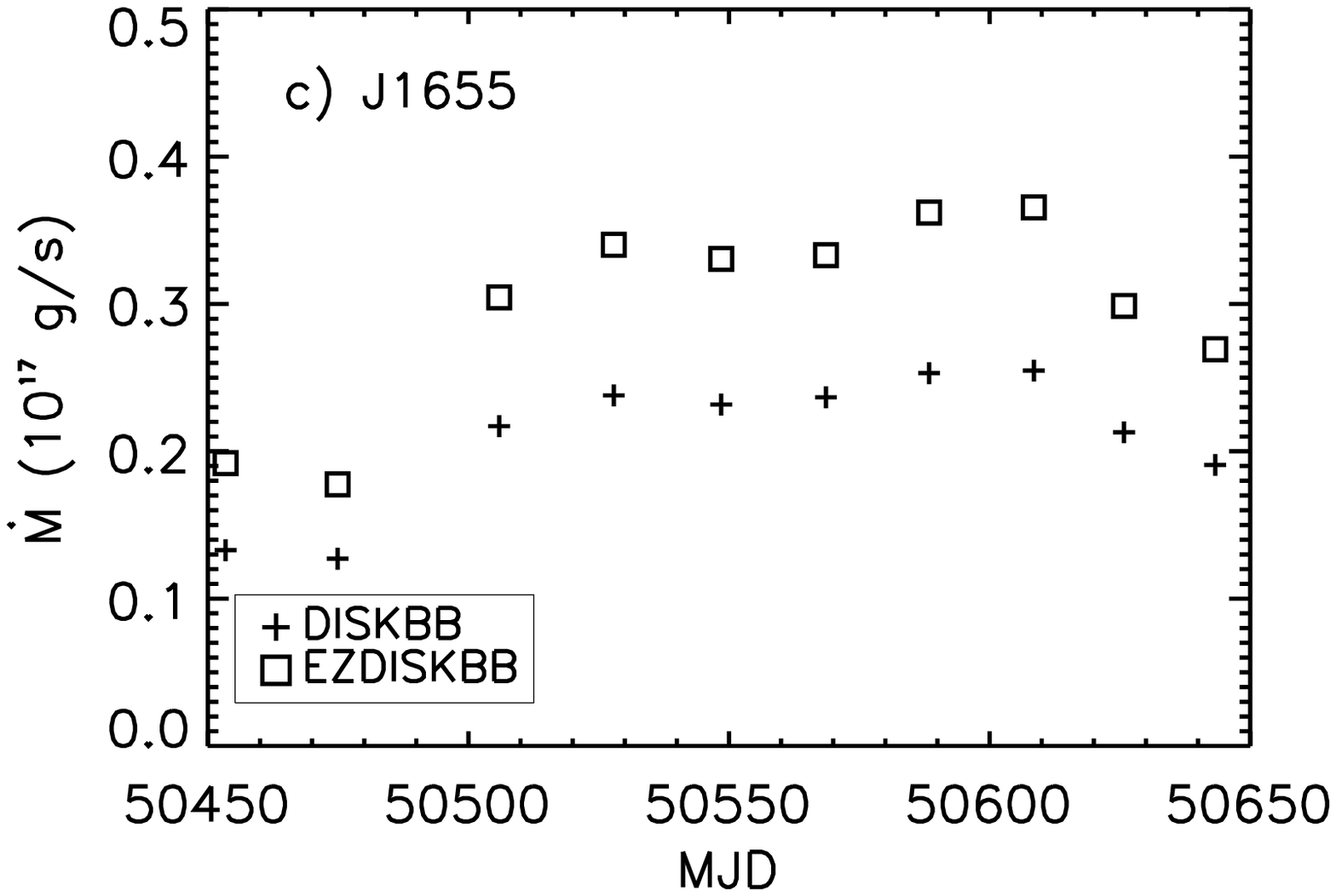}
\caption{\small{$\dot{M}$ versus time for the 10 spectra from each of
our BHXN, assuming a spectral hardening factor of $f=1.7$.  The plus
symbols indicate the values of $\dot{M}$ obtained from fits using
diskbb, and the squares indicate $\dot{M}$ values from fits using
ezdiskbb.  The vertical axes are in units of $10^{17}$ g/s.  We see
that ezdiskbb gives values for the accretion rate that are a factor of
$\approx$ 1.4 larger than those given by diskbb.  The ratio is
independent of $f$.}}
\end{figure}

We found that the systematic differences in the disk parameters for the
two models did not significantly affect the parameters of the power-law
component that we used in our fits.  Likewise, we found no systematic
differences between the parameters of the smedge and line components.
Finally, we compared the values of $\chi^2$ for our fits and found that
the quality of the fits was very similar for the two models.  Figure 7
shows the power-law parameters and $\chi^2$ versus time for U1543; the
other sources gave similar results.  Because the power-law (and other
minor) spectral parameters are scarcely affected when ezdiskbb is
substituted for diskbb, and also because the computed results (Table 4)
agree with the theoretical predictions, we are confident that the large
differences in $R_{\rm in}$ between the two models are genuine.

\begin{figure}
\epsscale{0.45}
\plotone{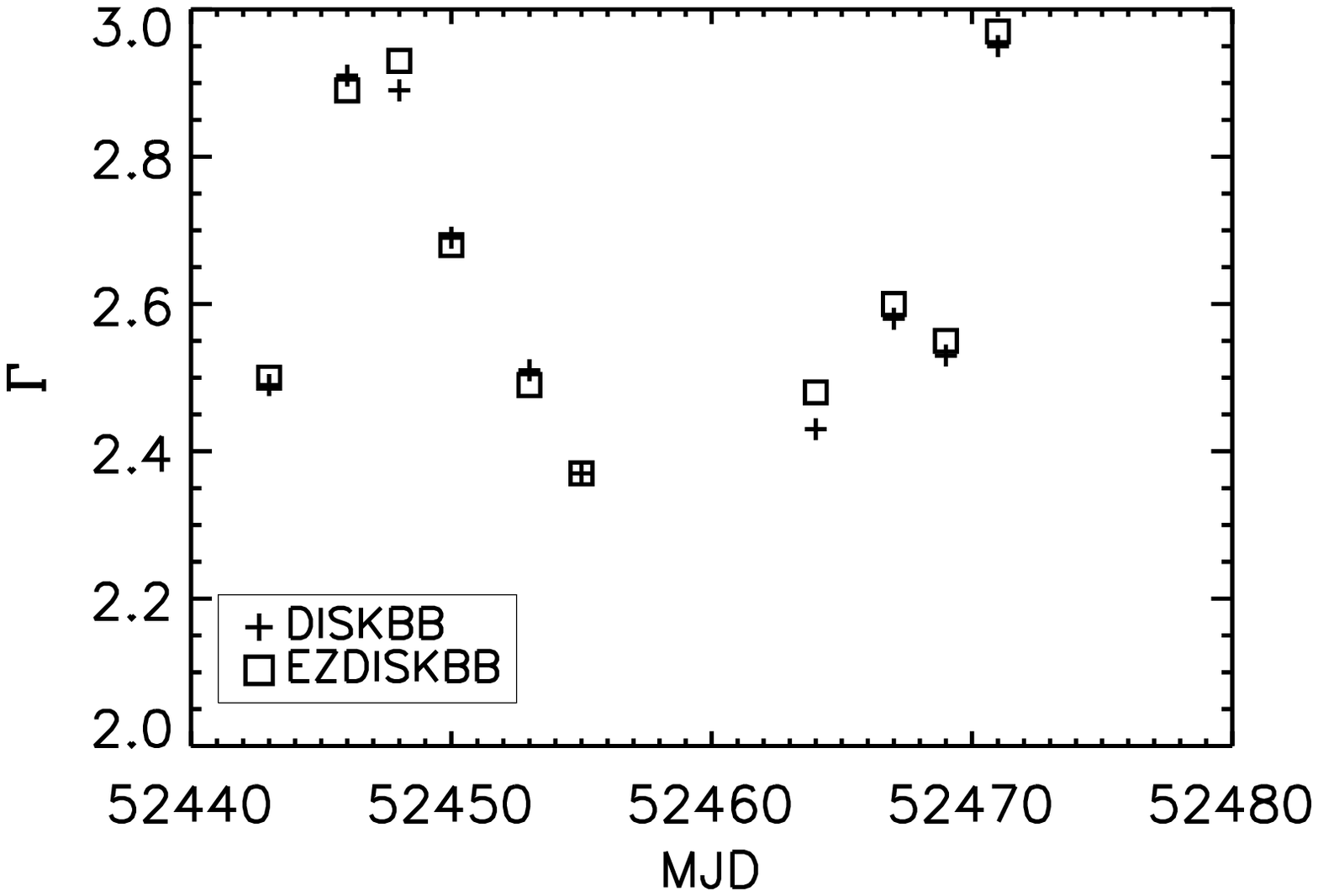}
\plotone{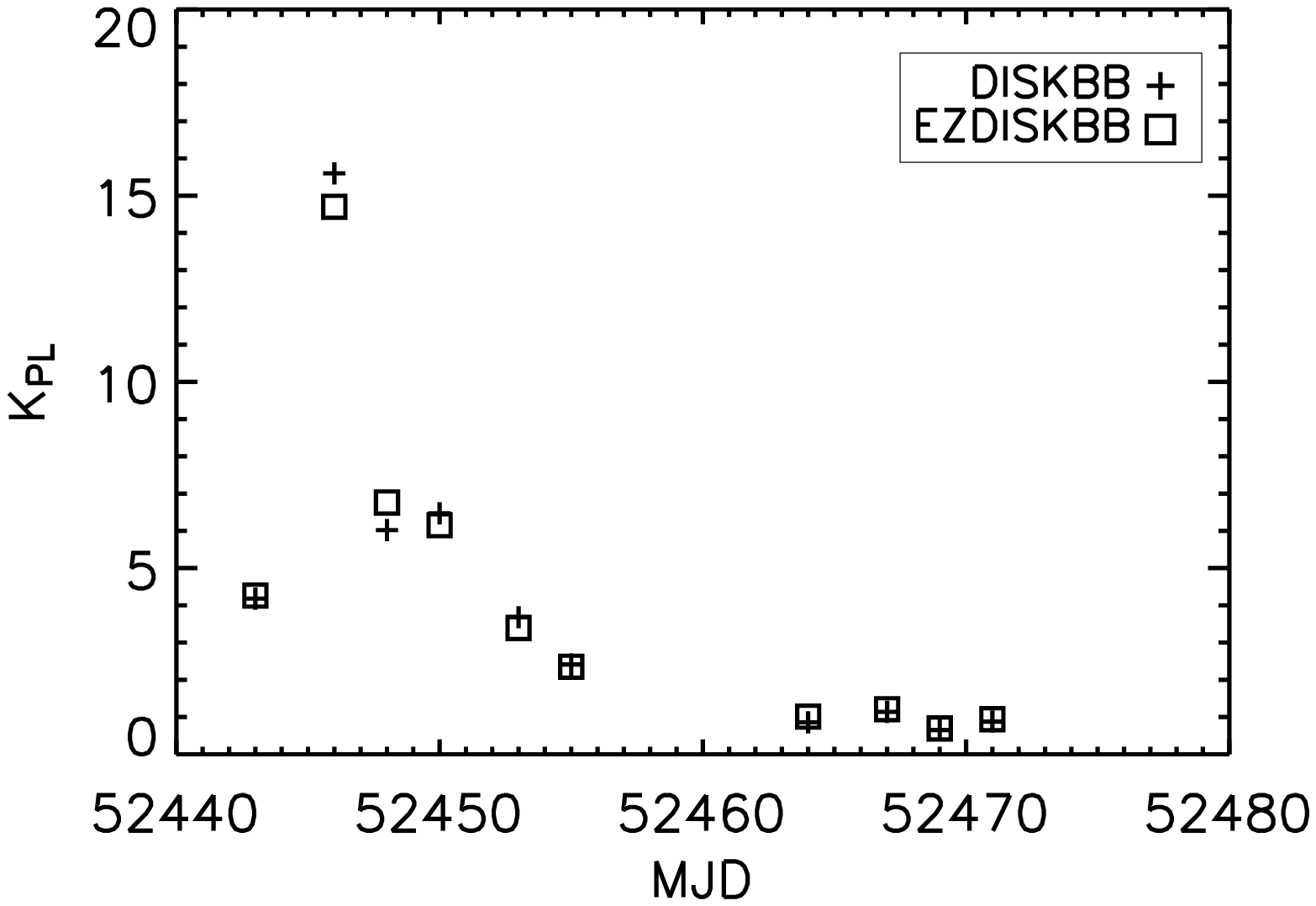}
\plotone{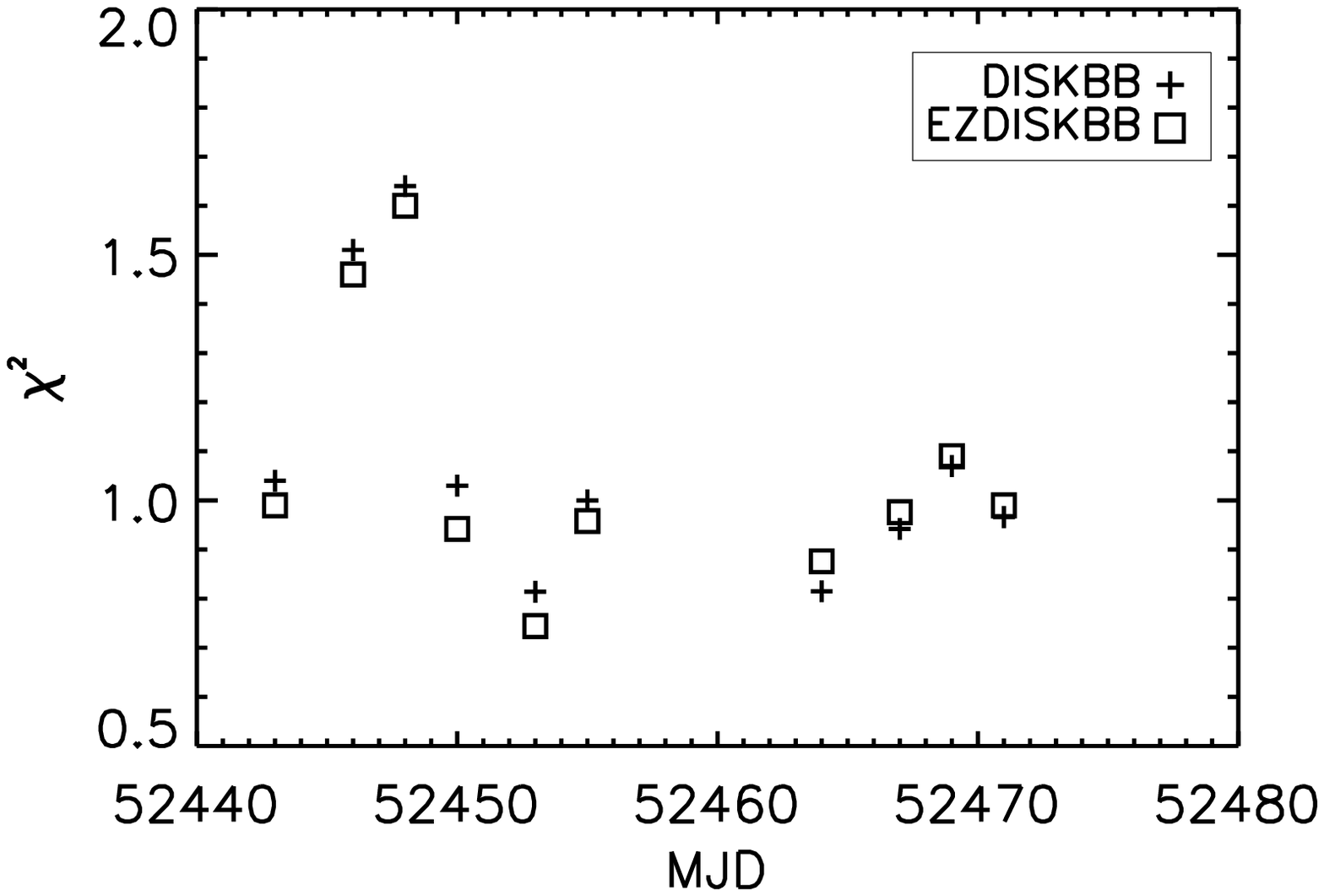}
\caption{\small{Power-law photon index ($\Gamma$), power-law
normalization ($K_{\rm PL}$), and $\chi^2$ versus time for U1543.  The
plus symbols indicate the values given by diskbb, while the squares
show the values given by ezdiskbb.  As the plots show, there were no
significant or systematic changes in the power-law component or the
quality of our fits when ezdiskbb was substituted for diskbb.}}
\end{figure}

\begin{deluxetable}{cccccccccc}
\tablecolumns{10}
\tablewidth{0pt}
\tabletypesize{\footnotesize}
\tablecaption{Diskbb Results vs. EZDiskbb Results\tablenotemark{a}}

\tablehead{
\multicolumn{1}{c}{} &
\multicolumn{3}{c}{$T_{\rm max}$} &
\multicolumn{3}{c}{$R_{\rm in}$} &
\multicolumn{3}{c}{$\dot{M}$}
	}

\startdata
 & Avg. & Max. & Min. & Avg. & Max. & Min. & Avg. & Max. & Min. \\
 & Ratio & Ratio & Ratio & Ratio & Ratio & Ratio & Ratio & Ratio & Ratio 
\\
 \hline

4U 1543-47 & 1.05 & 1.06 & 1.05 & 2.17 & 2.19 & 2.15 & 0.709 & 0.725 & 
0.699 \\
XTE J1550-564 & 1.05 & 1.06 & 1.05 & 2.19 & 2.21 & 2.17 & 0.714 & 0.735 
& 0.694 \\
GRO 1655-40 & 1.05 & 1.06 & 1.05 & 2.16 & 2.18 & 2.15 & 0.699 & 0.714 & 
0.690 \\
Overall & 1.05 & 1.06 & 1.05 & 2.17 & 2.21 & 2.15 & 0.709 & 0.735 & 
0.690 \\
\enddata

\tablenotetext{a}{The figures in the table correspond to the average,
maximum, and minimum of the ratios of the diskbb values to the
ezdiskbb values for each parameter and object.  The ratios are
independent of the spectral hardening factor $f$ assumed.}

\end{deluxetable}

\section{Discussion}

In the preceding section, we demonstrated that the zero-torque and
standard-torque boundary conditions imply systematically different
values for the disk parameters.  The change in $T_{\rm max}$ is not
very large (about 5\%), but the $T_{\rm max}$ value obtained with
diskbb is always higher than the value obtained with ezdiskbb by about
this amount.  The change in $\dot{M}$ is somewhat larger --- a
reduction of about 30\% --- and this difference again appears in every
spectrum.  The most significant change is in the value for the inner
radius of the disk: diskbb gives inner radii that are a factor of
$\approx$ 2.2 larger than those given by ezdiskbb.

This change in the size of $R_{\rm in}$ has important implications,
especially because the value of $R_{\rm in}$ for an accretion disk can
be used to estimate the angular momentum of the disk's central black
hole.  It has been shown that when a BHXN is in the high/soft state,
its inner radius remains remarkably stable over time (see Fig. 5) and
is therefore thought to be located near $R_{\rm ms}$ (for a review,
see McClintock \& Remillard 2004).  Measuring $R_{\rm in}$ therefore
allows us, in principle, to determine the approximate size of $R_{\rm
ms}$.  The size of $R_{\rm ms}$ depends on the dimensionless spin
parameter, $a_* \equiv a/R_{\rm g}$, where $a = J/cM$ ($J$ being the
BH angular momentum) and $R_{\rm g} \equiv GM/c^2$ (e.g., Shapiro \&
Teukolsky 1983).  The parameter $a_*$ can vary between $\simeq$ $-1$
and 1.  When $a_*$ = 0, we have a non-rotating (Schwarzschild) BH, and
$R_{\rm ms}$ = 6$R_{\rm g}$.  The maximum value of $a_*$ is less than
1 (Thorne 1974), but when $a_* \simeq$ 1, we have a maximally rotating
BH, and $R_{\rm ms} \simeq R_{\rm g}$.  When $a_* \simeq$ $-1$, the BH
is maximally rotating in the retrograde direction, and $R_{\rm ms}
\simeq 9R_{\rm g}$.  In between these extremes, $R_{\rm ms}$ decreases
monotonically as a function of increasing $a_*$ according to the
following formula \citep{bar72}:

\begin{eqnarray}
	R_{\rm ms}=R_{\rm g} \left\{ 3+Z_2 \mp [(3-Z_1)(3+Z_1+2Z_2)]^{1/2} 
\right\} , \\
	Z_1 \equiv 1+ \left(1- a_*^2 \right)^{1/3} \left[ \left(1+a_* 
\right)^{1/3}
	+ \left(1- a_* \right)^{1/3} \right] , \nonumber \\
	Z_2 \equiv \left( 3 a_*^2 + Z_1^2 \right)^{1/2} . \nonumber
\end{eqnarray}
In short, $R_{\rm ms}$ decreases monotonically with increasing $a_*$.
Since we have found that the zero-torque condition decreases our
values for $R_{\rm in}$ by a factor of 2.15 or more compared to the
standard-torque condition, our estimates for $a_*$ will increase
correspondingly by a large factor.

Figure 8 illustrates this effect by showing $R_{\rm ms}$ as a function
of $a_*$ for each of our BHXN, using the black hole masses listed in
Table 2.  Also shown is the average value of $R_{\rm in}$ for each
source.  We have plotted $R_{\rm in}$ as a shaded horizontal band in
order to account for the uncertainty in the value of the spectral
hardening factor, which we assume to be $f=1.7 \pm 0.2$ \citep{shi95}.
The band for each model therefore corresonds to the possible values of
the effective $R_{\rm in}$ when $f$ is between 1.5 and 1.9.  The band
with dashed lines corresponds to the $R_{\rm in}$ obtained with
ezdiskbb, and the band with dotted lines gives the inner radius for
diskbb.  Because we are assuming that $R_{\rm in} = R_{\rm ms}$, the
region in which the dashed band intercepts the curve for each black
hole gives the allowed range of values of $a_*$ for that black hole
using ezdiskbb.  Likewise, the intersection of the dotted band and the
curve gives the allowed range of $a_*$ for diskbb.

\begin{figure}
\epsscale{0.45}
\plotone{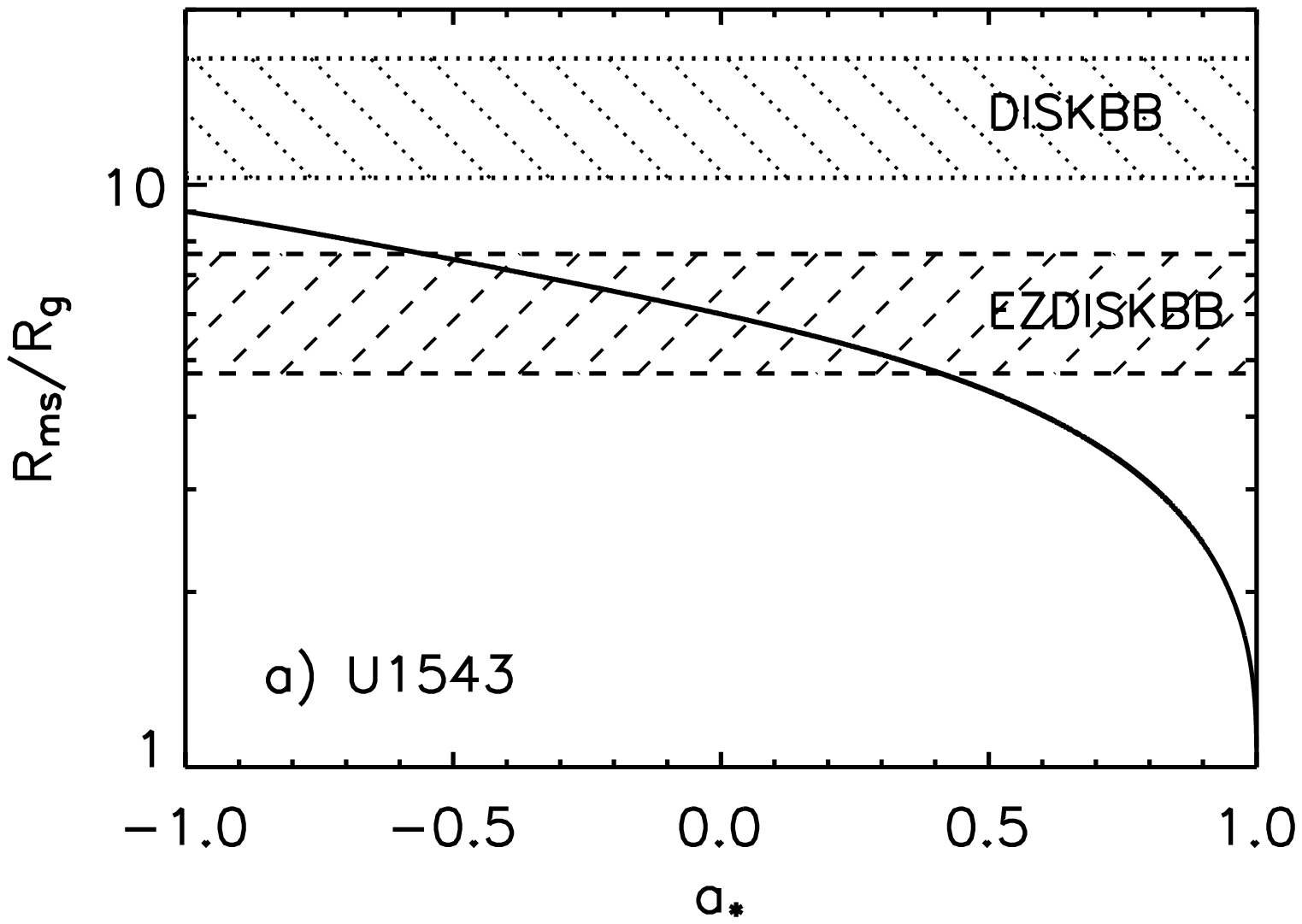}
\plotone{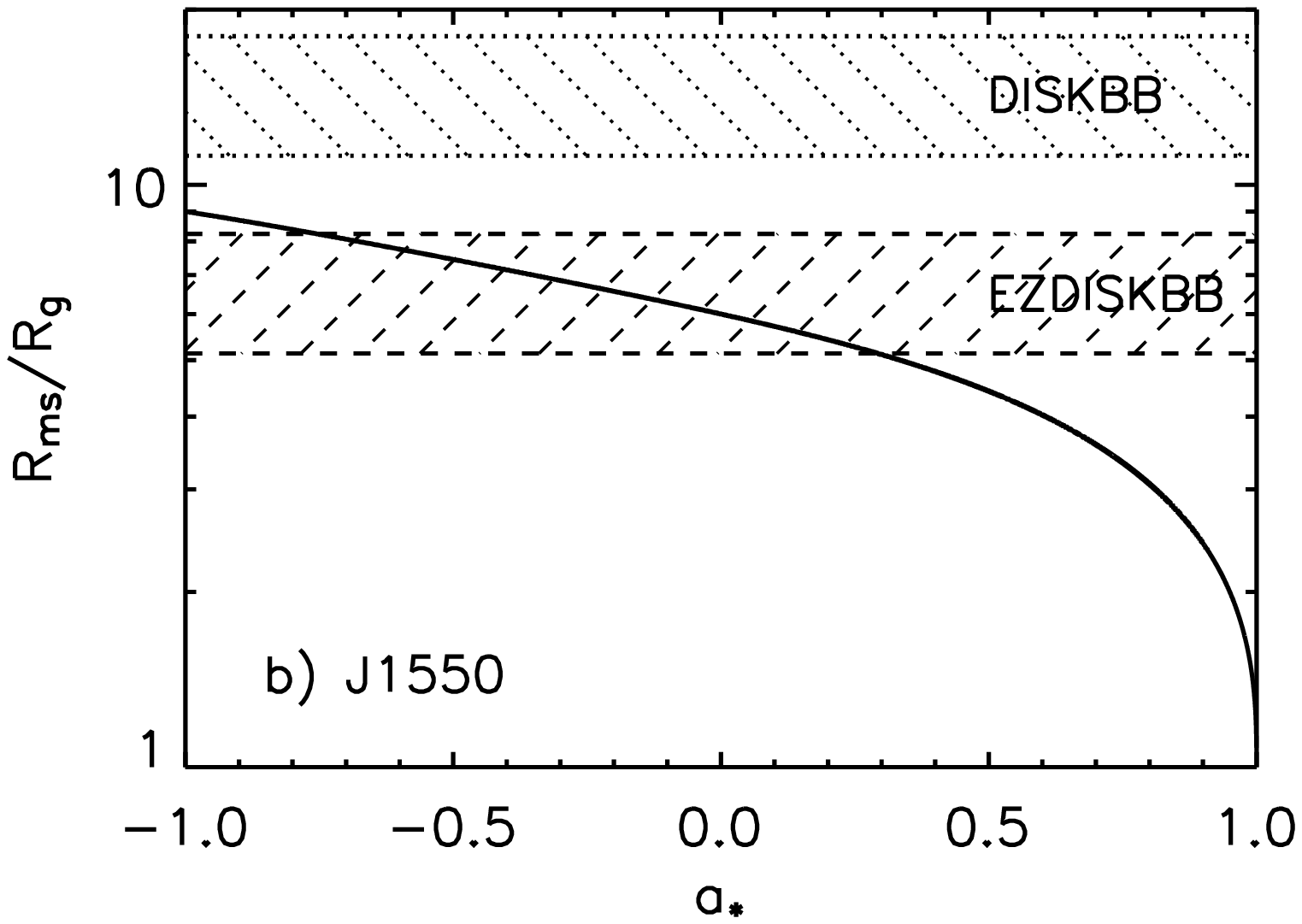}
\plotone{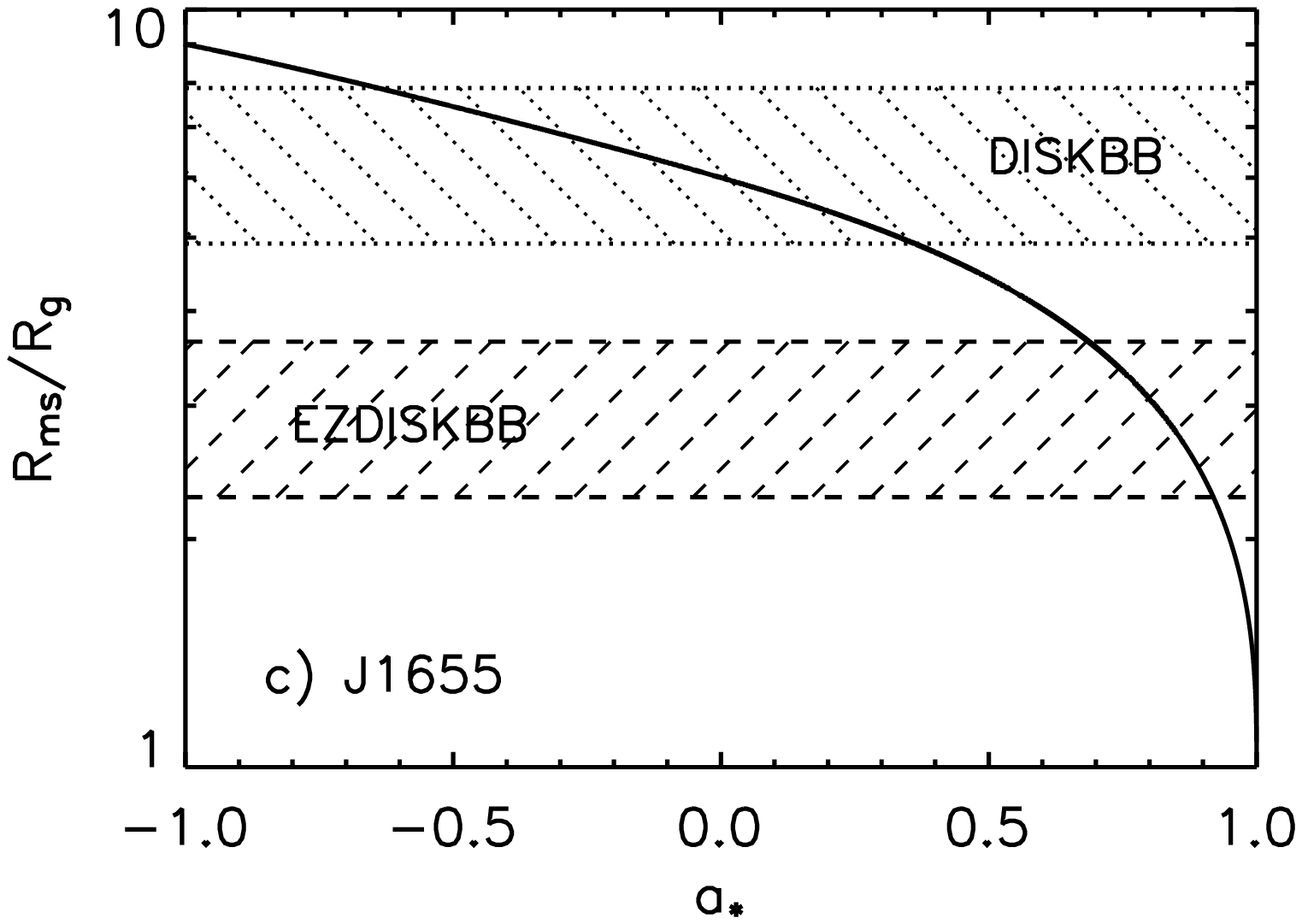}

\caption{Estimates of $a_*$ obtained using diskbb and ezdiskbb.  The
solid curves show $R_{\rm ms}$ as a function of $a_*$. The lower and
upper horizontal lines for each model correspond to the average values
of $R_{\rm in}$ when $f=1.5$ and $f=1.9$, respectively, which are the
limits on the spectral hardening factor according to Shimura \&
Takahara (1995).  The intersection of these lines and the curve gives
an estimate of the spin parameter of the black hole for that value of
$f$ and the model in question.  Ezdiskbb implies that U1543 and J1550
are slowly spinning, as the estimated values for $a_*$ are near
zero. Diskbb, on the other hand, gives values for the inner radius
that are too large to be consistent with any estimate for the spin
parameter.  For J1655, ezdiskbb indicates that $a_*$ should be between
about 0.7 and 0.9, while diskbb implies that $a_*$ is between $-0.7$
and 0.4.}
\end{figure}  

We can see immediately the implications of using a model that includes
the zero-torque boundary condition.  For U1543 and J1550, ezdiskbb
gives a range for the spin that is centered near zero, whereas diskbb
gives values for $R_{\rm in}$ that are too large to be consistent with
any allowed spin of these black holes.  For J1655, ezdiskbb implies a
rapidly spinning black hole with $a_*$ between about 0.7 and 0.9,
while diskbb gives an $a_*$ between about $-0.7$ and 0.4.  Sobczak
(2000) used diskbb, assuming $f=1.7 \pm 0.2$, to estimate an upper limit
on the spin of J1655 of about 0.7, but we see now that if we use
ezdiskbb, the spin can be greater than the Sobczak (2000) limit.  This
result for ezdiskbb supports past analyses that used high-frequency
quasiperiodic oscillations (QPOs) to infer high spin ($a_* \approx
0.93$) for J1655 \citep{cui98,str01}.  In summary, the choice of
boundary condition has a large impact on the inferred spin parameter
of the black holes that we have analyzed.  The zero-torque boundary
condition (ezdiskbb) implies substantially larger (prograde) spins
than the standard-torque boundary condition (diskbb).

Since the results vary dramatically for different assumed boundary
conditions, it is important to know which boundary condition is
appropriate for our systems.  As we discussed in \S 2, the zero-torque
assumption is valid either (i) when we have an accretion disk around a
slowly-spinning star, or (ii) when we have a thin accretion disk
around a black hole (or ultra-compact neutron star) with the inner
edge of the disk close to the radius of the marginally stable orbit.
In the black hole binaries that we have considered in this paper, the
latter assumption should hold in the very high and high/soft spectral
states, and perhaps also in some intermediate states.

When the accretion disk is truncated before it reaches the marginally
stable orbit, as in certain models of the low and quiescent state
(Narayan 1996; Esin, McClintock \& Narayan 1997; Esin et al. 1998, 2001),
the gas is thought to evaporate from the optically thick geometrically
thin disk to form an optically thin corona.  It is possible that, in
this case, the optically thick disk has a nonzero torque at its inner
edge.  However, there is at present no theoretical estimate of the
magnitude of the torque, so it is not clear how the system is to be
modeled.  Another case in which there might be a significant torque on
the inner edge is when a geometrically {\it thick} disk extends down
to the marginally stable orbit \citep{afs03}.  Indeed, MHD simulations
of thick advective disks by Hawley \& de Villiers (2004) do reveal
nonzero torques.  However, there is no simple prescription for the
magnitude of the torque, and there is no reason to believe that it
would match the standard torque assumed in diskbb.

In addition, when the torque on the inner edge is nonzero, the
accreting gas inside the inner edge presumably continues to experience
viscous angular momentum transport and energy dissipation (Gammie
1999; Agol \& Krolik 2000).  This additional source of heat will
clearly contribute to the observed radiation, but its contribution
will not be calculated in any thin disk model that stops the
computation at the inner edge.  Thus, a code like diskbb is
intrinsically incomplete: it implicitly assumes additional stresses
and dissipation inside the inner edge but it does not calculate the
corresponding emission.  In contrast, ezdiskbb is internally
consistent, at least for black hole accretion.  Since it assumes zero
torque at the inner edge, there is no additional source of radiation
inside the inner edge, and therefore the code does not make any error
by neglecting the gas inside $R_{\rm in}$.

The results shown in Figure 8 depend sensitively on the assumed value
of the spectral hardening parameter $f$.  In Figure~9 we show further
estimates of $a_*$ assuming no spectral hardening, i.e., $f=1$, which
is probably not a realistic assumption for any of the three sources
considered herein.  The values for $R_{\rm in}$ appear here as single
lines rather than bands because we are assuming a single value for
$f$.  In this case, our values for $R_{\rm in}$ decrease compared to
when $f=1.7$, and our estimated values for $a_*$ increase
correspondingly.  However, we still see a significant discrepancy
between ezdiskbb and diskbb.  Ezdiskbb now gives estimates of $a_*$
for U1543 and J1550 that are close to 0.9, meaning that these black
holes would be very rapidly spinning.  This estimate for the spin of
J1550 agrees well with independently estimated values of the spin that
were made using QPOs and iron lines \citep{mil01}.  Diskbb, however,
gives smaller estimates for the spins of U1543 and J1550 of about 0.5
and 0.3, respectively.  For J1655, ezdiskbb implies that $a_*$ is
almost equal to 1, while diskbb gives a slightly smaller value of
about 0.9.

\begin{figure}
\epsscale{0.45}
\plotone{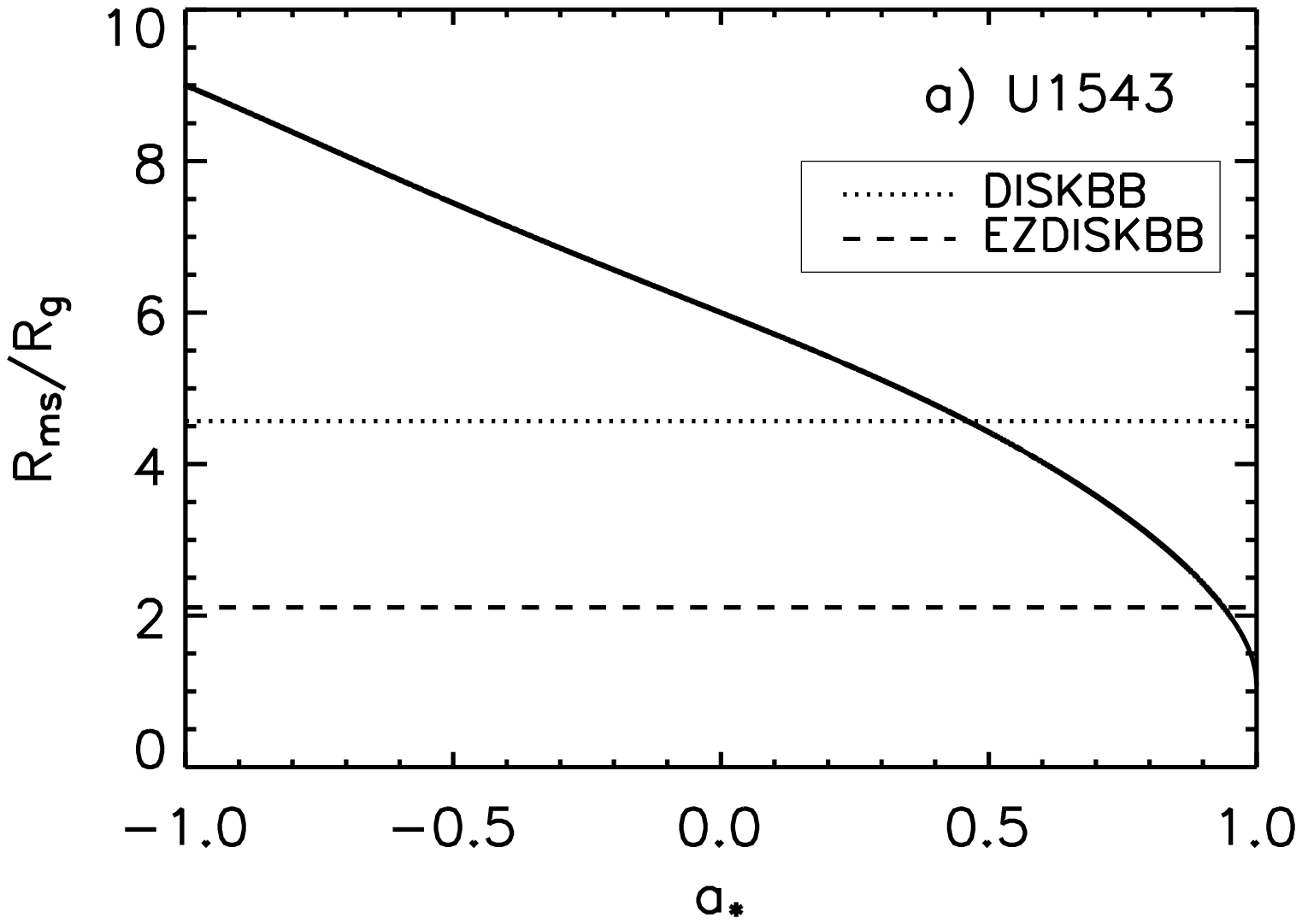}
\plotone{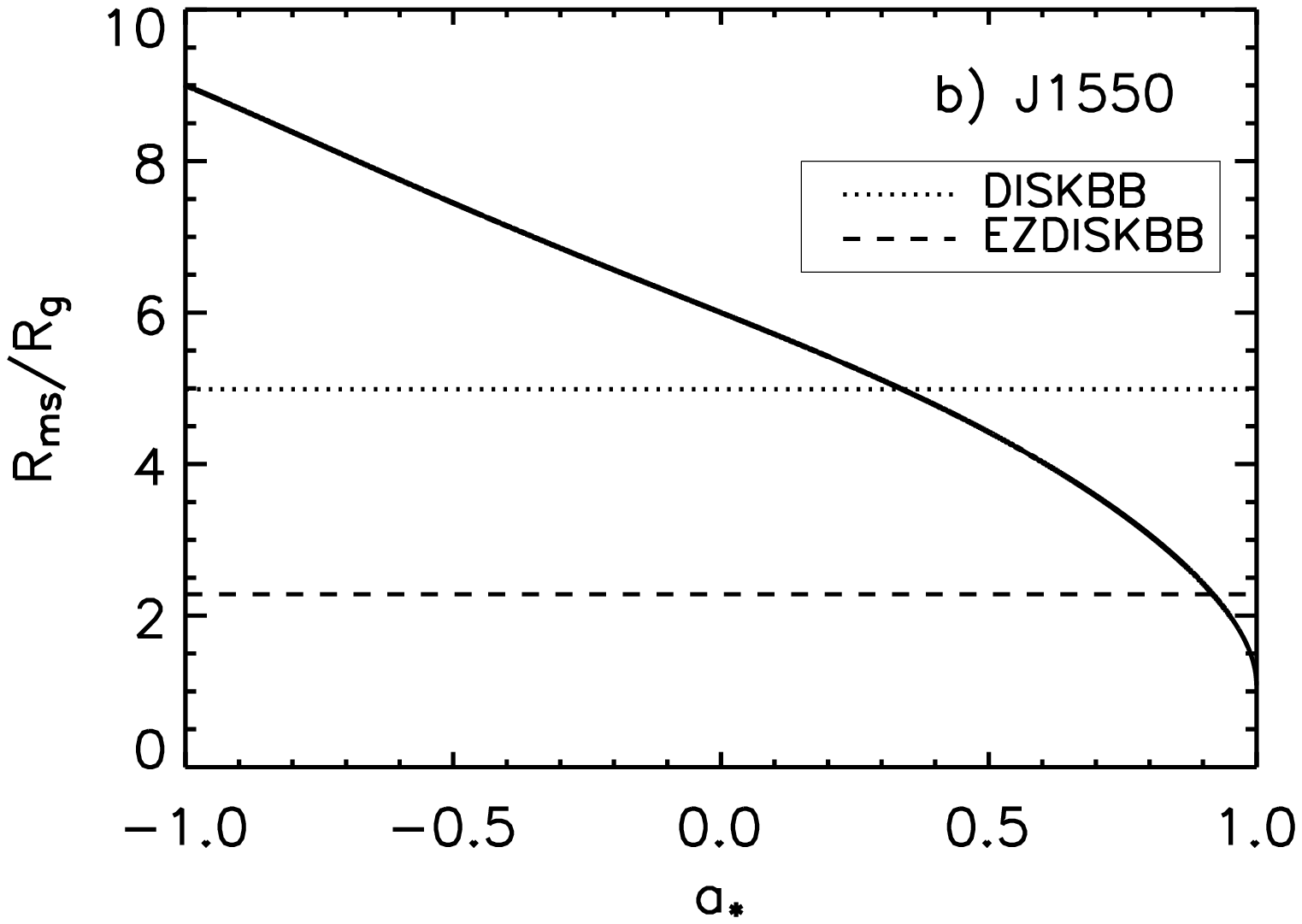}
\plotone{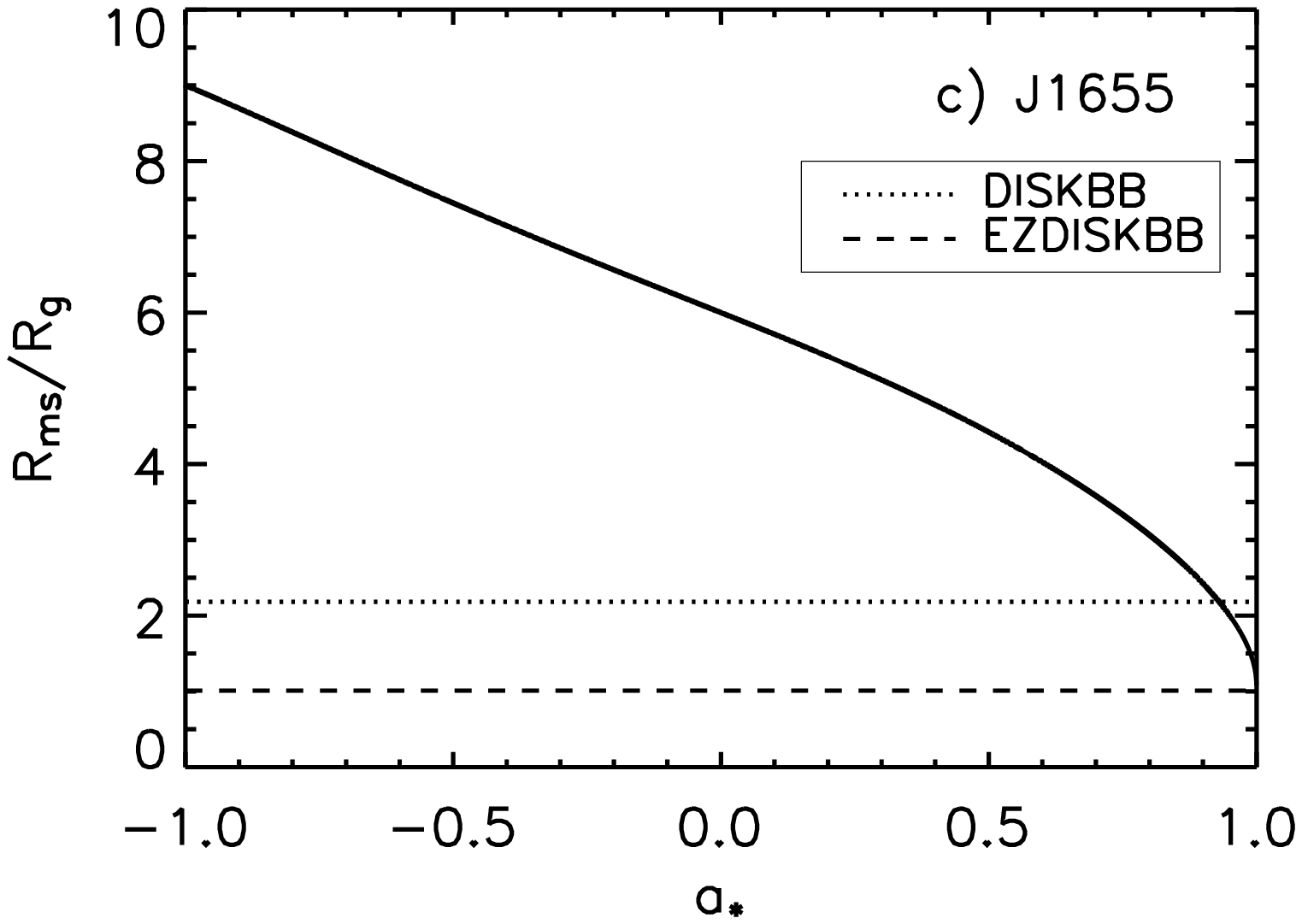}

\caption{Same as Figure 8 except that a spectral hardening factor of
$f=1$ is assumed, which implies that the values of $R_{\rm in}$ are
smaller and that each model corresponds to a single value of $R_{\rm
in}$.  We still see a large discrepancy between the values of $a_*$
inferred from ezdiskbb and diskbb.  Ezdiskbb implies an $a_*$ of about
0.9 for both U1543 and J1550, while diskbb gives estimates of about
0.5 for U1543 and 0.3 for J1550.  Using QPOs and Fe K lines, Miller et
al. (2001) have estimated that J1550 has high spin; ezdiskbb agrees
with this result much better than diskbb.  For J1655, both models
indicate high spin, with ezdiskbb giving an $a_*$ of almost 1 and
diskbb giving an $a_*$ of about 0.9. }
\end{figure}

What value of $f$ is appropriate for the systems we have analysed?  As
we mentioned in \S~2, the non-blackbody effects in the emitted
spectrum cannot be fully described by a single parameter $f$ --- in
reality, the modification to the blackbody emission is likely to be
different at each photon energy.  However, if we consider a simple
overall measure of the spectral modification, e.g., the shift in the
position of the spectral peak, then it is possible to describe the
modification crudely with a single number.  This is the philosophy
behind the use of $f$.  Shimura \& Takahara (1996) considered the
effects of Comptonization and concluded that, under conditions of
interest in X-ray astronomy, the change in the spectrum can be
described by shifting the temperature in the blackbody formula (which
determines the position of the peak) upward by $f\sim 1.5-1.9$
(average value $f\sim 1.7$), and at the same time changing the
normalization of the blackbody formula so as to keep the flux the
same.

Another relevant situation is when the escaping radiation is powered
by energy released deep inside the radiating gas at an optical depth
of several.  This might be relevant to optically thick accretion disks
if the viscous energy dissipation rate is proportional to the local
density or pressure of the gas.  If the gas has both absorptive and
scattering opacity, then the emerging radiation will have a color
temperature $T_{\rm color}$ larger than the effective temperature
$T_{\rm eff}$ (e.g., Rybicki \& Lightman 1979).  If the opacities are
independent of photon energy, then the shape of the spectrum continues
to be blackbody-like.  However, if the opacities vary with energy (the
usual case), then the spectral shape is also modified, as may be seen
in the neutron star spectra calculated by Zavlin et al. (1996).  For
simplicity, we might wish to ignore the change in the shape of the
spectrum and attempt to summarize the effects by a single parameter
$f$ that describes the shift in the peak of the spectrum.  As Figure 6
of McClintock et al. (2004) shows, the peak shifts by a factor of
$\sim1.65$, which is close to the value of 1.7 which we have used in
most of our analysis.

\section{Summary and Conclusions}

We have confirmed that diskbb, the very widely used MTB model in the
current version of XSPEC, assumes a nonzero torque at the inner boundary
of the accretion disk, which is not in accord with the classic and
current literature on thin-disk accretion.  We have therefore created a
new model, ezdiskbb, which assumes zero torque at the inner edge of the
disk.  We have fitted spectra from three well known BHXN with both
diskbb and ezdiskbb to compare the effect of using the different
boundary conditions, and we have found that diskbb gives values for the
inner radius of the disk in every spectral fit that are always a factor
of $\approx$ 2.2 greater than those given by ezdiskbb.  We have also
shown that a change in the size of the inner radius of the disk of this
magnitude has significant implications when estimating the spin
parameter of the disk's central black hole.

In many ways, ezdiskbb is functionally equivalent to the familiar MTB
model diskbb.  For example, both models are straighforward to use, are
very widely applicable, and their parameter sets are identical.  We
have clearly outlined ezdiskbb's underlying assumptions, and we have
tested it thoroughly.  The theoretical arguments for fitting with
ezdiskbb are based on the analysis of Afshordi \& Paczy\'nski (2003),
which indicates that the torque at the inner edge of the accretion
disk is small when the disk is thin and the inner edge of the disk is
close to $R_{\rm ms}$.  We therefore recommend that ezdiskbb be used
to analyze spectra for BHXN, especially in the high/soft state, as
well as in the very high, and some of the intermediate states, in
which the inner edge of the disk should be close to $R_{\rm ms}$.
However, there is still some debate about the magnitude of the torque
at the inner edge of the disk.  For example, Hawley \& De Villiers
(2004) argue that the torque in the inner disk region does not go to
zero, but their simulations correspond to a relatively geometrically
thick disk for which the Afshordi \& Pacsy\'nski (2003) argument does
not apply.  Because the question is still open, ezdiskbb provides a
useful means of comparison between the zero-torque and standard-torque
assumptions.  For ease and simplicity, results for the inner radius
obtained with diskbb can be divided by a factor of $\approx$ 2.2 in
order to estimate the inner radius that would be obtained using the
zero-torque inner boundary condition.  Similarly, mass accretion rate
estimates obtained with diskbb may be multiplied by $\approx 1.4$ to
obtain the corresponding rates for the zero-torque boundary condition.

The results obtained with either diskbb or ezdiskbb are quite
sensitive to the assumed value of the spectral hardening factor $f$.
We have presented arguments in \S~4 why the choice $f=1.7$ is
reasonable.  However, until more accurate disk atmosphere models
become available, this factor will represent a rather large
uncertainty in results derived from fitting data.

\acknowledgments

The authors would like to acknowledge K. Makishima for his forthcoming
and helpful discussions on diskbb.  They also acknowledge Andrzej
Zdiarski and Philip Kaaret for their important discussions, as well as
Ron Remillard for his assistance with the BHXN lightcurves, and Keith
Arnauld for his help with implementing ezdiskbb in XSPEC.  This work
has made use of the information and tools available at the HEASARC Web
site, operated by GSFC for NASA, and was supported in part by NASA
grants NAG5-9930 and NAG5-10780 and NSF grant AST 0307433.


\begin{thebibliography}

\bibitem[Afshordi \& Paczy\'nski(2003)]{afs03} Afshordi, N. \&
Paczy\'nski, B. 2003, \apj, 592, 354

\bibitem[Agol \& Krolik(2000)]{ago00} Agol, E. \& Krolik, J. H. 2000,
ApJ, 528, 161

\bibitem[Armitage, Reynolds, \& Chiang(2001)]{arm01} Armitage, P. J.,
Reynolds, C. S., \& Chiang, J. 2001, ApJ, 548, 868

\bibitem[Arnaud(1996)]{arn96} Arnaud, K. A. 1996, in ASP
Conf. Ser. 101: Astronomical Data Analysis and Systems V., 17

\bibitem[Bardeen et al.(1972)]{bar72} Bardeen, J. M., Press, W. H., \&
Teukolsky, S. A. 1972, ApJ, 178, 347

\bibitem[Cui, Zhang, \& Chen(1998)]{cui98} Cui, W., Zhang, S. N., \&
Chen, W. 1998, ApJ, 492, 53

\bibitem[Esin et al.(1997)]{esi97} Esin, A. A., McClintock, J. E,
\& Narayan, R. 1997, \apj, 489, 865

\bibitem[Esin et al.(1998)]{esi98} Esin, A. A., Narayan, R., Cui, W.,
Grove, J. E., \& Zhang, S.-N. 1998, \apj, 505, 854

\bibitem[Esin et al.(2001)]{esi01} Esin, A. A., McClintock, J.  E.,
Drake, J. J., Garcia, M. R., Haswell, C. A., Hynes, R. I., \& Muno,
M. P. 2001, \apj, 555, 483

\bibitem[Frank, King \& Raine(1992)]{fra92} Frank, J., King, A., \&
Raine, D.  1992, Accretion Power in Astrophysics (Cambridge University
Press)

\bibitem[Gammie(1999)]{gam99} Gammie, C. F. 1999, \apj, 522, L57

\bibitem[Gierli\'nski et al.(1999)]{gie99} Gierlinski, M., Zdziarski,
A. A., Poutanen, J., Coppi, P.  S., Ebisawa, K., \& Johnson,
W. N. 1999, \mnras, 309, 496

\bibitem[Gierli\'nski \& Done(2004)]{gie04} Gierlinski, M. \& Done,
C. 2004, MNRAS, 347, 885

\bibitem[Hawley \& De Villiers(2004)]{haw04} Hawley, J. F. \& De
Villiers, J. P. 2004, to be published in Progress of Theoretical
Physics Supplement, astro-ph/0402665

\bibitem[Hjellming \& Rupen(1995)]{hje95} Hjellming, R. M. \& Rupen,
M. P. 1995, Nature, 375, 464

\bibitem[Jahoda et al.(1996)]{jah96} Jahoda, K., Swank, J. H., Giles,
A. B., et al. 1996, in Proc. SPIE, Vol. 2808, EUV, X-ray, and
Gamma-ray Instrumentation for Astronomy VII, ed. O. H. Siegmund \&
M. A. Gummin, 59

\bibitem[Kluzniak, W.(1987)]{klu87} Kluzniak, W. 1987, Ph.D. thesis,
Stanford University

\bibitem[McClintock et al.(2004)]{mcc04b} McClintock, J. E., Narayan,
R., \& Rybicki, G. B. 2004, \apj, in press (astro-ph/0403251)


\bibitem[McClintock \& Remillard(2004)]{mcc04} McClintock, J. E. \&
Remillard, R.  A. 2004, astro-ph/0306213

\bibitem[Makishima et al.(1986)]{mak86} Makishima, K., Maejima, Y.,
Mitsuda, K., et al.  1986, \apj, 308, 635

\bibitem[Miller et al.(2001)]{mil01} Miller, J. M., Wijnands, R.,
Homan, J., et al. 2001, ApJ, 563, 928

\bibitem[Mitsuda et al.(1984)]{mit84} Mitsuda, K., Inoue, H., Koyama,
K., et al. 1984, \pasj, 36, 741

\bibitem[Narayan(1996)]{nar96} Narayan, R. 1996, \apj, 462, 136

\bibitem[Novikov \& Thorne(1973)]{nov73} Novikov, I. D. \& Thorne,
K. S. 1973, Black Holes - Les Astres Occlus, ed. C. D. Witt \&
B. S. D. Witt (Gordon \& Breach), 343

\bibitem[Orosz \& Bailyn(1997)]{oro97} Orosz, J. A. \& Bailyn,
C. D. 1997, \apj, 477, 876

\bibitem[Orosz et al.(1998)]{oro98} Orosz, J. A., Jain, R. K., Bailyn,
C. D., McClintock, J. E., \& Remillard, R. A. 1998, \apj, 499, 375

\bibitem[Orosz et al.(2002)]{oro02} Orosz, J. A., Groot, P. J., van
der Klis, M. et al. 2002, ApJ, 568, 845

\bibitem[Orosz(2003)]{oro03} Orosz, J. A. 2003, in preparation

\bibitem[Paczy\'nski(1991)]{pac91} Paczy\'nski, B. 1991, \apj, 370,
597

\bibitem[Park et al.(2003)]{par03} Park, S. Q., Miller, J. M.,
McClintock, J. E., et al. 2003, accepted for publication in ApJ,
astro-ph/0308363

\bibitem[Popham \& Narayan(1991)]{pop91} Popham, R. \& Narayan,
R. 1991, ApJ, 370, 604

\bibitem[Popham \& Narayan(1995)]{pop95} Popham, R. \& Narayan,
R. 1995, ApJ, 442, 337

\bibitem[Pringle \& Rees(1972)]{pri72} Pringle, J. E. \& Rees,
M. J. 1972, \aap, 21, 1

\bibitem[Pringle(1981)]{pri81} Pringle, J. E. 1981, \araa, 19, 137

\bibitem[Reynolds \& Armitage(2001)]{rey01} Reynolds, C. S. \&
Armitage, P. J. 2001, ApJ, 561, L81

\bibitem[Rutledge et al.(1999)]{rut99} Rutledge, R. E., Bildsten, L.,
Brown, E. F., Pavlov, G. G., \& Zavlin, V. E. 1999, \apj, 514, 945

\bibitem[Rybicki \& Lightman(1979)]{ryb79} Rybicki, G. B., \&
Lightman, A. P. 1979, Radiative Processes in Astrophysics, New York:
Wiley

\bibitem[Shakura \& Sunyaev(1973)]{sha73} Shakura, N. I. \& Sunyaev,
R. A. 1973, \aap, 24, 337

\bibitem[Shapiro \& Teukolsky(1983)]{sha83} Shapiro, S. L. \&
Teukolsky, S. A. 1983, Black Holes, White Dwarfs, and Neutron Stars
(John Wiley \& Sons)

\bibitem[Shimura \& Takahara(1995)]{shi95} Shimura, T. \& Takahara,
R. 1995, ApJ, 445, 780

\bibitem[Sobczak et al.(1999)]{sob99} Sobczak, G. J., McClintock,
J. E., Remillard, R. A., Bailyn, C. D., \& Orosz, J. A. 1999, ApJ,
520, 776

\bibitem[Sobczak(2000)]{sob00a} Sobczak, G. J. 2000, PhD thesis,
Department of Astronomy, Harvard University

\bibitem[Sobczak et al.(2000)]{sob00b} Sobczak, G. J., McClintock,
J. E., Remillard, R. A., et al.  2000, ApJ, 544, 993

\bibitem[Strohmayer(2001)]{str01} Strohmayer, T. E. 2001, ApJ, 552,
L49

\bibitem[Thorne(1974)]{tho74} Thorne, K. S. 1974, ApJ, 191, 507

\bibitem[Zavlin et al.(1996)]{zav96} Zavlin, V. E., Pavlov, G. G., \&
Shibanov, Yu. A. 1996, A\&A, 315, 141

\bibitem[Zhang, Cui, \& Chen(1997)]{zha97} Zhang, S. N., Cui, W., \&
Chen, W. 1997, ApJ, 482, L155


\end{thebibliography}
\end{document}